# *The Effect of Chemical Disorder on Defect Formation and Migration in Disordered MAX Phases*

Prashant Singh,[a*] Daniel Sauceda,[a] and Raymundo Arroyave [a]

[a]*Department of Materials Science & Engineering, Texas A&M University, College Station, TX 77843, USA*

## Abstract

MAX phases have attracted increased attention due to their unique combination of ceramic and metallic properties. Point-defects are known to play a vital role in the structural, electronic and transport properties of alloys in general and this system in particular. As some MAX phases have been shown to be stable in non-stoichiometric compositions, it is likely that such alloying effects will affect the behavior of lattice point defects. This problem, however, remains relatively unexplored. In this work, we investigate the alloying effect on the structural-stability, energy-stability, electronic-structure, and diffusion barrier of point defects in MAX phase alloys within a first-principles density functional theory framework. The vacancy ($V_M$, $V_A$, $V_X$) and antisite (M-A; M-X) defects are considered with M and A site disorder in $(Zr\text{-}M)_2(AA')C$, where M=Cr,Nb,Ti and AA'=Al, Al-Sn, Pb-Bi. Our calculations suggest that the chemical disorder helps lower the $V_A$ formation energies compared to $V_M$ and $V_X$. The $V_A$ diffusion barrier is also significantly reduced for M-site disorder compared to their ordered counterpart. This is very important finding because reduced barrier height will ease the Al diffusion at high-operating temperatures, which will help the formation of passivating oxide layer (i.e., $Al_2O_3$ in aluminum-based MAX phases) and will slow down or stop the material degradation. We believe that our study will provide a fundamental understanding and an approach to tailor the key properties that can lead to the discovery of new MAX phases.

## Introduction

Research in high-temperature structural materials is gaining momentum due to the ongoing quest for ever increasing operating efficiencies in power generation technologies [1]. An example is the nuclear power industry, which seeks new reactor designs that in turn demand structural and coating materials capable of withstanding higher operating temperatures at higher irradiation conditions. MAX phases are a new class of high-temperature materials with the general formula $M_{n+1}AX_n$ [n=1,3; M=transition element; A=A-group, and X=C or N]. Their high-temperature thermodynamic and structural stability, damage-tolerance and oxidation-resistance [2-5] make MAX-phases a potential candidate structural material for nuclear applications [6]. Beyond structural and coating applications, research on MAX phases has also led to the discovery of MXenes based on the selective leaching of the A element [7,8]. Many MAX phase (and MXene) properties of interest in these materials depend on the nature and behavior of point defects within the lattice.

In the past, irradiation has been shown to significantly alter the microstructures of systems through the creation of point defects [9-12]. For example, in the context of nuclear applications, irradiation degrades the material by damage accumulation. These defects interact with other defects and atoms within the solid-solution while they migrate through the lattice [13]. It is thus to be expected that understanding the behavior of (extrinsic or intrinsic) point defects in MAX phases is important if one is to understand the factors controlling their performance under irradiation environments. In many materials systems, chemical disorder-induced alloy complexity is another factor that can significantly impact the defect evolution, energy dissipation, and radiation resistance [14-17] that arises from the interaction of (energetic) ions with solids in a radiation environment.

Following the trends observed in the development of other materials systems, recent efforts on the synthesis, characterization and modeling of MAX phases has focused on the expansion of the composition palette through mixing in the different sublattices of the MAX phase crystal system. While some work has been carried out already in understanding the energetics and behavior of point defects in stoichiometric MAX phases [12], the investigation of defect physics of disordered MAX phases remains relatively unexplored relative to other materials systems. It becomes desirable to understand the interplay of disorder and the point defects to develop an understanding towards the microstructural changes due to irradiation for the future nuclear structural materials [18-27]. Recent studies show that disorder can be used to tailor alloy properties to achieve desirable physical and chemical properties, e.g., thermoelectric-effects [28], charge-transport [29], and magnetoresistance [30]. The tailoring of properties in these applications ultimately arises from the strong connection between chemical disorder and the stability of point defects in those materials systems. It is thus to be expected that, in a similar fashion, chemical disorder in MAX phase crystal systems plays an important role in determining the local configuration, formation energetics and migration behavior of point defects in these systems.

In this work, we discuss the impact of chemical disorder on the behavior of point defects of MAX phases. Five MAX phase solid solution systems [based on the stoichiometry $M_2AlC$, M=Ti, Zr, Nb, Cr, Ta (for Ta see supplement)], of the 211 stoichiometry, are chosen to investigate the connection between alloying and point defect behavior. We choose to investigate the effect of single-site (M-, or A-), and two-site (M- & A-) disorder on the energetics and migration behavior of point defects. Our calculations show, for example, that modifying chemical disorder leads to notable changes on the defect formation and migration energies. Our focus remains on Al-containing systems, which arises from their superior oxidation resistance that results from the formation (in some of these compounds) of stable alumina-based passivating layers that protect them from further oxidation [31]. Specifically, we apply first-principles density functional theory and climbing nudge elastic band (cNEB) schemes to study $V_M$, $V_A$ and $V_C$ vacancy-mediated diffusion in $(Zr\text{-}M)_2AlC$ and $(Zr\text{-}M)_2(AA')C$, where M=Cr,Nb,Ti and A-A'=Al, Al-Sn, Pb-Bi. Zr-containing MAX phases were selected because they have already been identified as candidate materials for nuclear structural applications. The detailed discussion about the dependence of the *migration energy* on the alloying elements and defects is provided. We also investigate the responsible electronic structure features that control the properties of defects and its connection to elemental properties, such as, atomic-radii and electronegativity.

**Computational details**

While in the context of Density Functional Theory (DFT) there are many approaches used nowadays to account for chemical disorder in a crystal structure, the use of periodic supercells that mimic disordered configurations constitutes the most common method to compute the electronic and structural properties of disorder materials with and without defects [32]. In this work, we use the of special quasi-random supercell structures (SQS) to model chemically-disordered MAX phase supercells as implemented within the Alloy Theoretic Automated Toolkit (ATAT) [33]. SQS are periodic unit cells in which atomic occupancy is engineered to match as close as possible the statistics of truly random alloys at the same chemical composition. The quality of the generated SQS (periodically repeated) supercell is quantified by the correlation function (we choose two and three body interactions to achieve best correlation function).

In this work, an SQS of 200 atoms has been generated to mimic disorder on M-, A- and/or X-site sublattices. Both cationic and anionic vacancy defects are considered in the disordered configurations $(M\text{-}M')_2(A\text{-}A')(X\text{-}X')$. Antisite pairs are created by interchanging the M (A) or A (M) atoms within neighboring layers. We note here that while smaller cells could have been used, the selected supercell size of $O(10^2)$ is necessary in order to accurately compute the energetics of periodic structures without the pathologies arising from their periodic images [34,35].

We use first-principles density functional theory as implemented in Vienna *Ab-initio* Simulation Package (VASP) [36,37] to study structural properties, electronic properties and vacancy diffusion in disorder MAX phases. The gamma-centered Monkhorst-Pack [38] k-mesh of 1x1x1 and 3x3x3 is used for Brillouin zone integration during geometry-optimization and charge self-consistency, respectively. Total energies and forces are converged to $10^{-5}$ eV/cell and -0.001 eV/Å. The Perdew, Burke and Ernzerhof (PBE) generalized gradient approximation is used with a planewave cut-off energy of 533 eV [39].

The vacancy formation energy ($E_{form}^{Vac}$) in MAX phases is defined as:

$$E_{form}^{Vac}\ (V_{M/A/X}) = E_{tot}\ (V_{M/A/X}) - E_{tot}(MAX) + \mu_{M/A/X}$$

where M/A/X, M (=Ti, Ta, Cr, Nb, Zr), A=Al, X=C. $E_{tot}$ ($V_{M/A/X}$) is the calculated total energy of a cell with defect and $E_{tot}$(MAX) is the total energy of a pure MAX phase without defects, and $\mu_{M/A/X}$ is the chemical potential of $V_{M/A/X}$. For the calculation of defect energies, we use the chemical potential of an isolated gas-phase atom and consider the vacancy defects as charge-neutral due to metallic nature of MAX phases. The chemical potential of isolated atoms is calculated in the same MAX phase SQS supercell by keeping M/A/X atoms at the center of the unit-cell. For this work, we decided not to include the entropic effect because the energy scales (for the formation of defects and migration energies) tend to be in the order of 100 meV and above, with some processes requiring energies well above 4 eV. Therefore, the major conclusions are unlikely to change.

The defect migration (or diffusion barrier) energies are calculated using the cNEB method [40] as implemented in VASP. The cNEB method can be used to identify the minimum energy path of migrating defects from one site to another site within the same unit cell by essentially identifying the set of configurations along the path with vanishingly small atomic forces normal to the path itself. Thus, the cNEB method allows us to understand plausible mechanisms for defect migration as well as the energy barriers associated to these atomic long-range displacements. This information is important not only to understand the kinetics of defect evolution but can also contribute to better understanding and identification of plausible atomic mechanisms leading to defect recovery [41]. Here, the vacancy migration energies in ordered and disordered MAX phase crystals is determined by a transition state search, where the initial and final states are relaxed structures with one vacancy, respectively. We use 6 intermediate images to accurately calculate the migration energy barrier of MAX phases [42] from lowest energy position to the other symmetry equivalent position.

## Results and discussion

**Structural properties and formation energies of ordered MAX phases with and without vacancies**: $M_2AlC$ phase forms a hexagonal structure (space-group=$P6_3/mmc$) that consists of edge-sharing $M_6X$ (M=cation, X=anion) octahedra interleaved with cation (A)-layer. The octahedra is similar to the rock salt binary carbides. While our interest is on disordered MAX phases, we first present structural and energy stability of ordered MAX phase and vacancy cases. **Table I** shows the ground-state structural properties and effect of vacancy on $M_2AlC$, where M=Cr,Ta,Nb,Ti,Zr. We compute the equilibrium lattice configurations of each case and find good agreement with experimentally known MAX phases [43-46], ensuring the reliability and accuracy of our calculations. We allow the vacancy supercell to relax homogeneously to the ground state energies before evaluating the complete set of the elastic constants. In **Table. S1**, we list the elastic constants ($C_{ij}$), bulk modulus (K), shear modulus (G), Young's modulus I, Poisson ratio ($\nu$) and Pugh ratio k= G/K as obtained by Voigt-Reuss-Hill approximation [47]. The lower Pugh (k) ratio for some MAX phases indicates relative brittleness compared to other chemistries. Comparing the bulk moduli of $Zr_2AlC$ with $Nb_2AlC$ from **Table S1** shows that $Nb_2AlC$ is elastically much stiffer than usually known allotropes. This is because $Nb_2AlC$ has larger elastic moduli all across: a bulk modulus of 170 GPa, a shear modulus of 119 GPa and a Young's modulus of 290 GPa compared to $Zr_2AlC$.

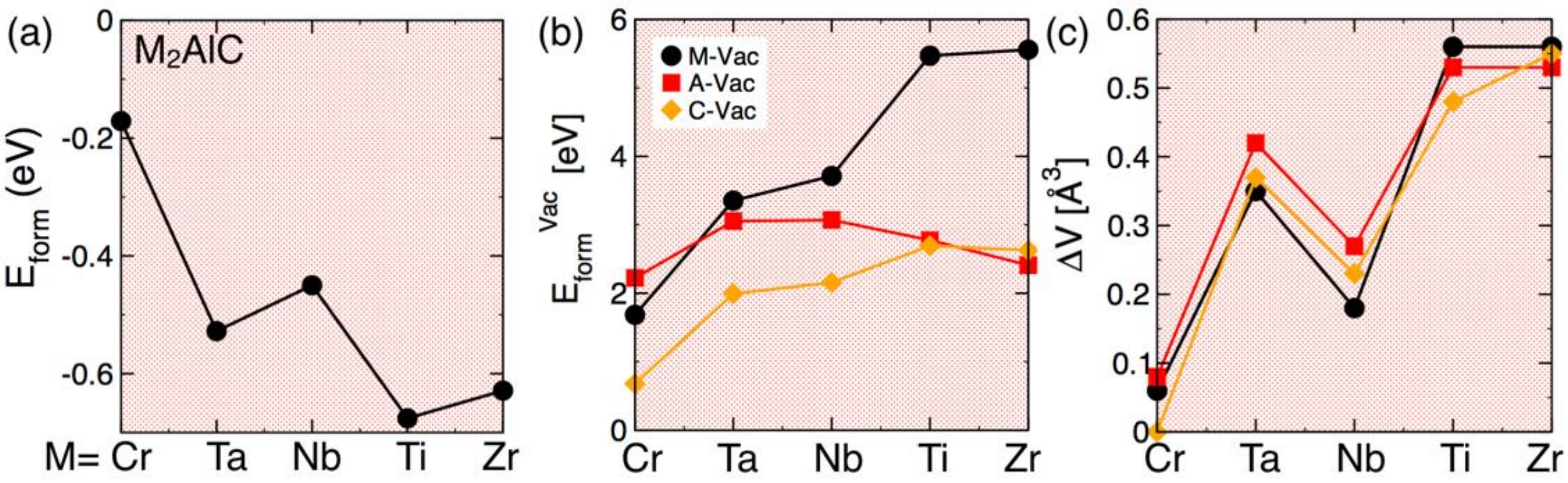

**Figure 1**. The formation enthalpy of (a) pure and (b) vacancy ($V_M$,$V_A$,$V_X$) $M_2AlC$ MAX phases, M=Cr,Ta,Nb,Ti,Zr. (c) Volume change of vacancy MAX phase with respect no-vacancy cases.

In **Fig. 1**, we show (a) the formation enthalpy of ordered MAX phases, (b) defect stability ($E_{form}^{Vac}$), and (c) relative change in cell volume [$\Delta V = V_{final}(M_2AlC\text{-vacancy}) - V_{initial}(M_2AlC)$] between pure and defect-containing structures. The formation energy ($E_{form}$) in **Fig. 1a** suggests that Ti- and Zr-based MAX alloys are more stable—relative to their elemental constituents—than their other transition metal counterparts. The large $V_M$ and $V_C$ vacancy formation energies in **Fig. 1b** makes it harder to create M-& C-vacancies for $Ti_2AlC$ and $Zr_2AlC$. On other hand, the A-site vacancy formation is less energetically favorable in $Nb_2AlC$ and $Ta_2AlC$. The volume of the MAX phase, in **Fig. 1c**, is sensitive to the vacancy with positive $\Delta V$. The $Cr_2AlC$ with $V_C$ is the only exception in **Fig. 1c** that shows either very small or no change in volume (attributed to magnetic character of Cr). If we make a comparison, the stable phases in **Fig. 1b** have smaller shear and Young's moduli (see **Table. S1)**. The Poisson's ratio of most of the ordered and vacancy MAX phases lie within the usual MAX phase range, i.e., from 0.200 to 0.260. The Pugh ratio (k) for crystalline $Ti_2AlC$ (1.29) is smaller than other MAX phases, while some vacancy-containing alloys in **Table S1** are fairly large. This indicates that those phases are relatively close to the ductile regime—here we note that the ability of MAX phases to undergo significant deformation is the result of complex mechanisms that require analysis well beyond the relatively simplistic metrics of bulk and shear modulus and that remain outside the scope of this work.

A major focus of our work is to look at the effect of chemical alloying on the formation energy, vacancy stability and vacancy migration of disordered MAX phases. As mentioned above, the fact that MAX phases can be synthesized with solid solutions on the M, A, or X sites greatly increases their chemical versatility and significantly expands the MAX phase composition space. In order to emulate disordered solid solutions, we designed an SQS supercell with 50:50 disorder on M-site, and M-&A-site without vacancy to test the $E_{form}$ compared to their ordered MAX phase counterparts. The calculated $E_{form}$ of $(Zr\text{-}M)_2AlC$ and (Zr-M)2(A-A')C are shown in **Fig. 2** where M=Cr, Nb, Ti and A-A' =Sn-Al, Pb-Bi. The horizonal lines in **Fig. 2** represents $E_{form}$ of the order phase. For example, alloying M-site in $Nb_2AlC$ with Zr stabilizes the $(ZrNb)_2AlC$ by −0.2 eV/atom, however, if we think of alloying the M-site in $Zr_2AlC$ with Nb this slightly reduces the stability by + 0.05 eV/atom. This indicates the tendency of C−Nb−Al to order while C−Zr−Al tends to cluster compared to their ordered counterpart. The overall picture of negative formation energy shows the thermodynamic stability of each of disordered MAX phase, relative to their elemental constituents, considered in this study. Our simulations also suggest that the compositional dependence of the energy of mixing is complex and non-ideal [48]. The minimum in the energy of mixing obtained for ordered (**Fig. 1**) and disordered (**Fig. 2**) MAX phases is consistent with the existence of a solubility limit.

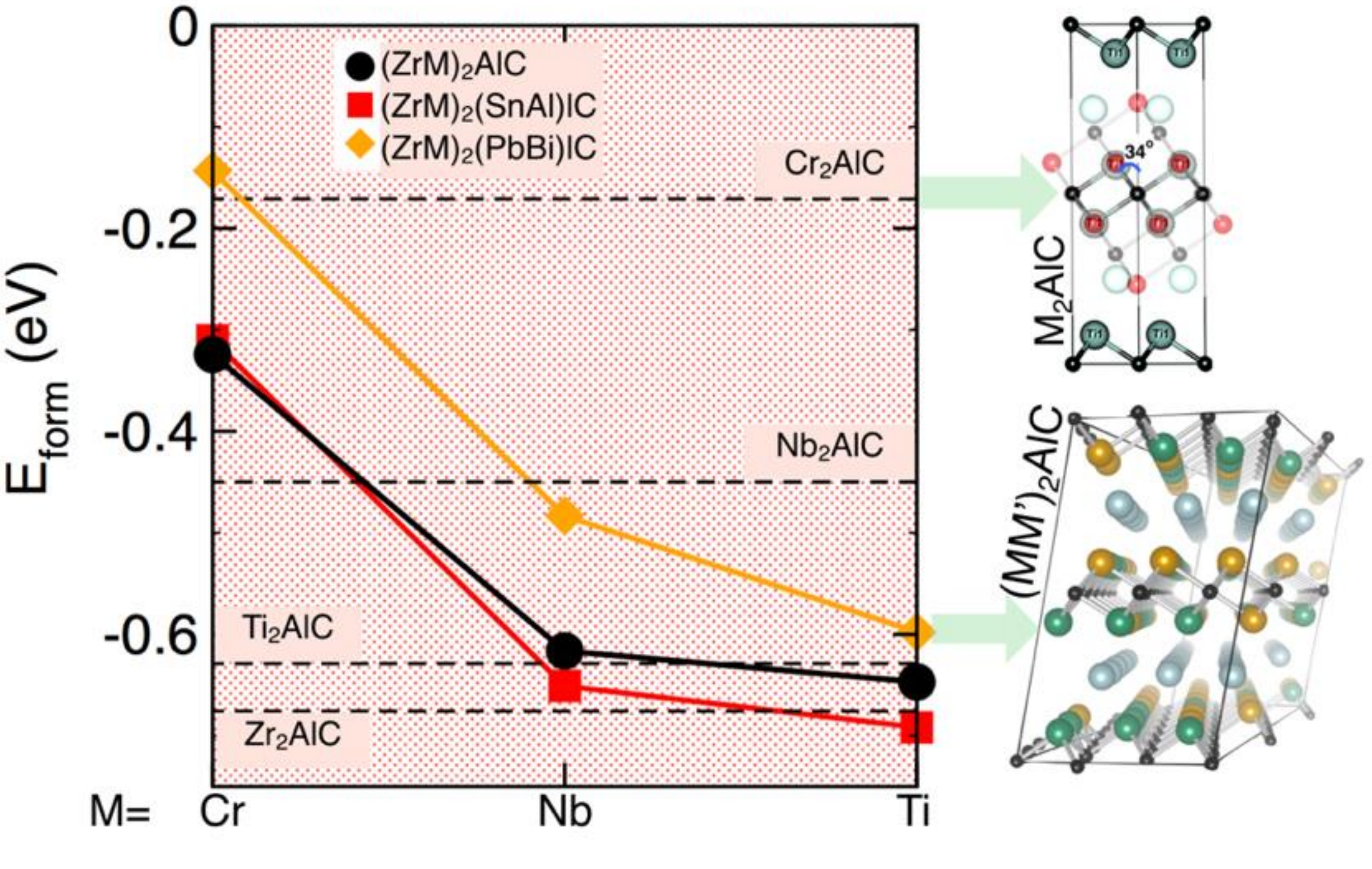

**Figure 2**. Formation enthalpy ($E_{form}$) of MAX phase for M-site disorder and horizonal dashed (black) lines represent order MAX-phase $E_{form}$. Right-panel shows (top) ordered, $M_2AlC$, and (bottom) disordered $(M\text{-}M')_2AlC$ supercell. Note that comparison with formation energies of pure MAX phases is just qualitative as we have not computed the convex hull against all possible phases in these systems.

**Vacancy defects:** In this section, we discuss the effect of vacancies on volume and $E_{form}^{Vac}$ of pure and disordered max phases. We investigate three types of vacancies, i.e., $V_M$, $V_A$ and $V_X$. **Fig.**

**3 (a,c,e)** shows the trends of positive volume change with respect to no-vacancy disorder max phase. Point-defecst can lead to overall change in volume with respect to ideal cell, e.g., the defects in nuclear graphite induced by irradiation increases the $c_{lat}$ and decreases ($a_{lat}$ & $b_{lat}$) in [0001] plane in response to interstitial carbons. The introduction of vacancies increases the volume as shown in **Fig. 1c & 3**. The present study shows that M-elements from group-VI of the periodic table, i.e. Cr or Nb based MAX phase are the only exceptions in that they exhibit a drop in relative volume with respect to other cases. Two site-disordered $(ZrNb)_2(SnAl)C$ MAX phase shows negative $\Delta V$. The change in volume is very small but a significant drop in $E_{form}^{Vac}$is found compared to one-site disorder. The calculated $E_{form}^{Vac}$ in **Fig. 3(b,d,f**) shows that $V_M$ in $(ZrTi)_2AlC$ are harder to be formed compared to $(ZrNb)_2AlC$ and $(ZrCr)_2AlC$, i.e., $E_{form}^{Vac}$ [$(ZrTi)_2AlC$] > $E_{form}^{Vac}$ [$(ZrNb)_2$ AlC] > $E_{form}^{Vac}$ [$(ZrCr)_2$ AlC]. However, the $E_{form}^{Vac}$ is lower for two site-disordered $(ZrM)_2[AA']C$, where M=Cr,Nb,Ti and AA'= Sn-Al, Pb-Bi, i.e., vacancy creation is easier compared to the single-site disordered case. The $(ZrCr)_2(SnAl)C$ is the only exception, where two site disorder increases $V_M$ and $V_A$ $E_{form}^{Vac}$compared to $(ZrNb)_2(SnAl)C$ and $(ZrTi)_2(SnAl)C$. On the other hand, [$(ZrCr)_2(SnAl)C$] and [$(ZrNb)_2(SnAl)C$] have competing $E_{form}^{Vac}$ as shown **Fig. 3d**. The $E_{form}^{Vac}$ trend in **Fig. 3f** for [$(ZrM)_2(PbBi)C$] is significantly reduced compared to **Fig. 3b**: $E_{form}^{Vac}$ [$(ZrTi)_2(PbBi)C$] > $E_{form}^{Vac}$ [$(ZrNb)_2(PbBi)C$] > $E_{form}^{Vac}$ [$(ZrCr)_2(PbBi)C$], i.e., it is easier to create $V_A$ vacancies compared $V_X$ in $(ZrCr)_2AlC$.

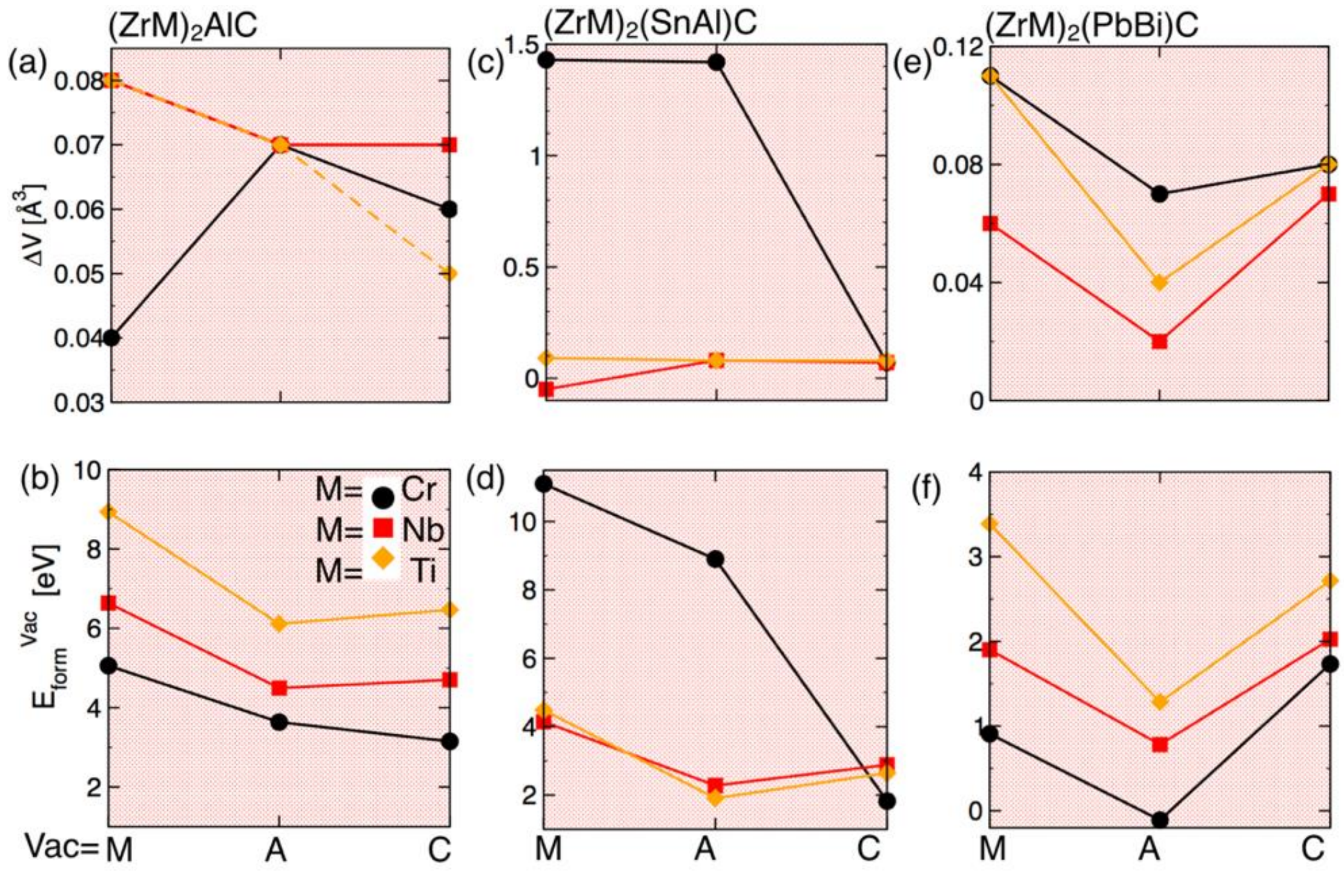

**Figure 3**. The effect of vacancy on volume and vacancy formation energy of disorder MAX phase. We consider M-and A-site disorder along with three possible vacancy scenarios: $V_M$, $V_A$ and $V_X$. We consider $(Zr\text{-}M)_2AlC$ and $(Zr\text{-}M)_2(A\text{-}A')C$ cases where M=Cr, Nb, & Ti, and A-A'=Sn-Al & Pb-Bi, respectively.

For M=Cr, Nb and Ti in $(Zr\text{-}M)_2AlC$, the $E_{form}^{Vac}$ approximately follows the trend [$E(V_M) > E(V_A) > E(V_X)$], [$E(V_M) > E(V_X) > E(V_A)$] and $E(V_M) > E(V_X) > E(V_A)$ for M=Ti, respectively. Notably, $E_{form}^{Vac}$ depends strongly on choice of atomic $\mu_{M/A/X}$. This is the reason, $E_{form}^{Vac}$ for $V_A$, and/or $V_C$ is lower in $(Zr\text{-}M)_2AlC$ than $(Zr\text{-}M)_2(SnAl)C$ or $[(Zr\text{-}M)_2(PbBi)C$ as shown **Fig. 3d & 3f**. The $V_{Zr}$ is the least affected while $V_C$ is the most affected defect in disordered MAX phases due to the respective atomic $\mu_{Zr/C}$. Therefore, irrespective of the choice of μ in ordered MAX phase, it is difficult to create $V_{Zr}$ as shown in **Fig. 1b**. Depending on choice of disorder, $V_{Zr}$ becomes relatively easier to form, while it is harder to create $V_A$ compared to $V_X$ in $Zr_2AlC$. Consequently, the synthesis condition, i.e., choice of μ, decides the relative stability of $V_{Al}$ or $V_C$ in $Zr_2AlC$. Contrastingly, choice of μ on other MAX phases has a negligible effect on the $E_{form}^{Vac}$, where $V_C$ remain the most stable defect configurations.

**Antisite defect**: The $E_{form}^{antisite}$ indicates the difficulty of antisite defect formation and irradiation incited recovery mechanism, which depends on the vacancy types and the target sublattices. The effect of antisite defects on volume (ΔV), energy stability and charge density difference of pure and defected MAX phases are shown in **Fig. 4a** & **4b** for $(Zr\text{-}M)_2AlC$. The $E_{form}^{antisite}$ do not have significant dependence on the atomic configuration as the atomic species are only interchanged, i.e., no atoms are moved in-or-out of the chemical reservoir. We show in **Fig. 4a** that M-Al antisite pair does not affect ΔV, but the exchange between M and X increases the volume – with Ti-C as only exception that shows negative ΔV. Antisite defects leads to positive ΔV for $(ZrCr)_2AlC$ while negative ΔV for $(ZrCr)_2AlC$ and $(ZrTi)_2AlC$. Based on $E_{form}^{Antisite}$ in **Fig. 4b**, it is harder to create antisite pairs at Zr (Zr-C or Zr-Al). Comparison of $E_{form}^{Antisite}$ (**Fig. 4b**) with $E_{form}^{Vac}$ (**Fig. 3b**) suggests that antisite defects for M=Cr are favorable as they have lower formation energies.

We can understand this in terms of empirical parameters, such as, electronegativity (χ) and atomi-radii (R). Here, the electronegativity difference of two atoms can tell us about the displacement of charge when two chemical species are mixed together. For example, in $(ZrCr)_2AlC$, $\Delta\chi_{Cr\text{-}Al}$ is less than $\Delta\chi_{Zr\text{-}Al}$ or $\Delta\chi_{Zr\text{-}Al}$ for $\chi_{Zr} = 1.33$, $\chi_{Cr}= 1.66$, $\chi_{Al} = 1.61$, and $\chi_C = 2.5$ and ΔR of Cr-Al is less than Zr-C or Al-C for $R_{Zr} = 0.86$ Å, $R_{Cr} = 0.76$ Å, $R_{Al} = 0.53$ Å, and $R_C = 0.29$ Å, respectively. The large difference in ΔR (62% vs 43%) and Δχ (21% vs 3%)) while comparing Zr-Al with Cr-Al, we see ΔR and Δχ in $(ZrCr)_2AlC$ explains large formation energy of the Zr-Al pair. However, Δχ and ΔR suggest that Cr-Al is the preferred pair of antisite defects where electronic effects play a pivotal role. Clearly, the C-layer is not energetically favorable for Zr/Cr or Al for antisite defect formation, therefore, it creates a vacancy by migrating into the next available cation M- (or A) layer. This shows that interchanging M- and A-cation position with X-anion costs more energy than anion-anion interchange.

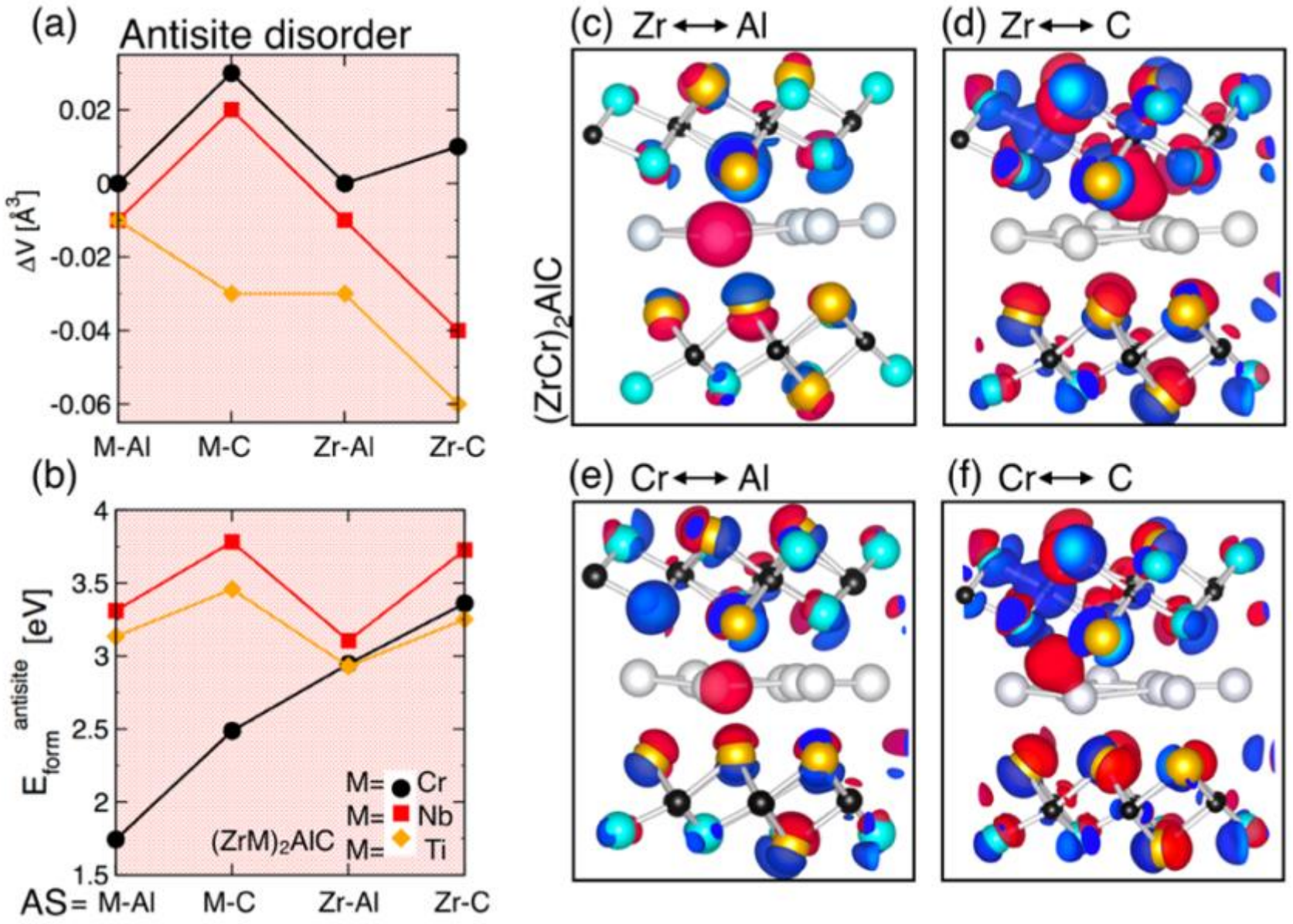

**Figure 4**. Effect of antisite defects on (a) cell volume and (b) formation energies of $(ZrM)_2AlC$, where M=Cr,Nb,TI. (c-f) Charge density difference between $(ZrM)_2AlC$ antisite defect and disorder cells.

The charge density difference between (Zr-M)$_2$AlC antisite defect and disordered cells in **Fig. 4e** shows that the in-plane Zr-*d*/Cr-*d* interactions shows large charge transfer for M= Cr. The Zr-Cr/Zr-Zr/Cr-Cr bonds in $(ZrCr)_2AlC$ are populated with in-plane Zr-Cr/Zr-Zr/Cr-Cr bonds, respectively, whereas the in-plane Zr-Cr/Zr-Zr/Cr-Cr bonds are emptied in favor of the Zr-Cr/Zr-Zr/Cr-Cr basal bonds. This can help us to understand the reason behind the increased stability of the Zr/Cr-C layers in the $(ZrCr)_2AlC$ system. To exemplify, we plot $Ti_2AlC$ and $Ta_2AlC$ band structure in **Fig. S7** that shows metallic behavior in the basal plane with multiple bands crossing at the Fermi-level ($E_{Fermi}$) while no bands cross the $E_{Fermi}$ in c-direction along ($\Gamma$-A).

**Electronic properties**: In **Fig. 5**, we show the total density of state of disordered (Zr-M)$_2$AlC MAX phases for M=Cr, Nb and Ti. The transition elements Zr, Cr, Nb and Ti have partially filled bonding d-states (also see **Fig. S8**), which means that the electrons at $E_{Fermi}$ are mainly from M elements and form conducting bands. If we look at the partial DOS of the $Zr_2AlC$ and $Cr_2AlC$ in **Fig. S8,** the Zr-*d* and Cr-*d* states in (-5.5 eV to -2.5 eV) and (-8.5 eV to -5 eV) energy range, respectively, overlap significantly with the C-*p* states. Addition of Cr to $Zr_2AlC$ at the Zr site add more electrons that increase the DOS near the $E_{Fermi}$ (**Fig. 5**), i.e., Cr addition increases the localized d-states. This also signifies that Zr-*d*/Cr-*d* & C-*p* in the alloy form strong covalent bonds, while, the Al-*p* with Zr-*d* and Cr-*d* states in the energy range -3eV to 0 eV and -5 eV to -2 eV show a weaker hybridization, respectively. This is suggesting metallic bonding between Al-*p* and M-*d* states.

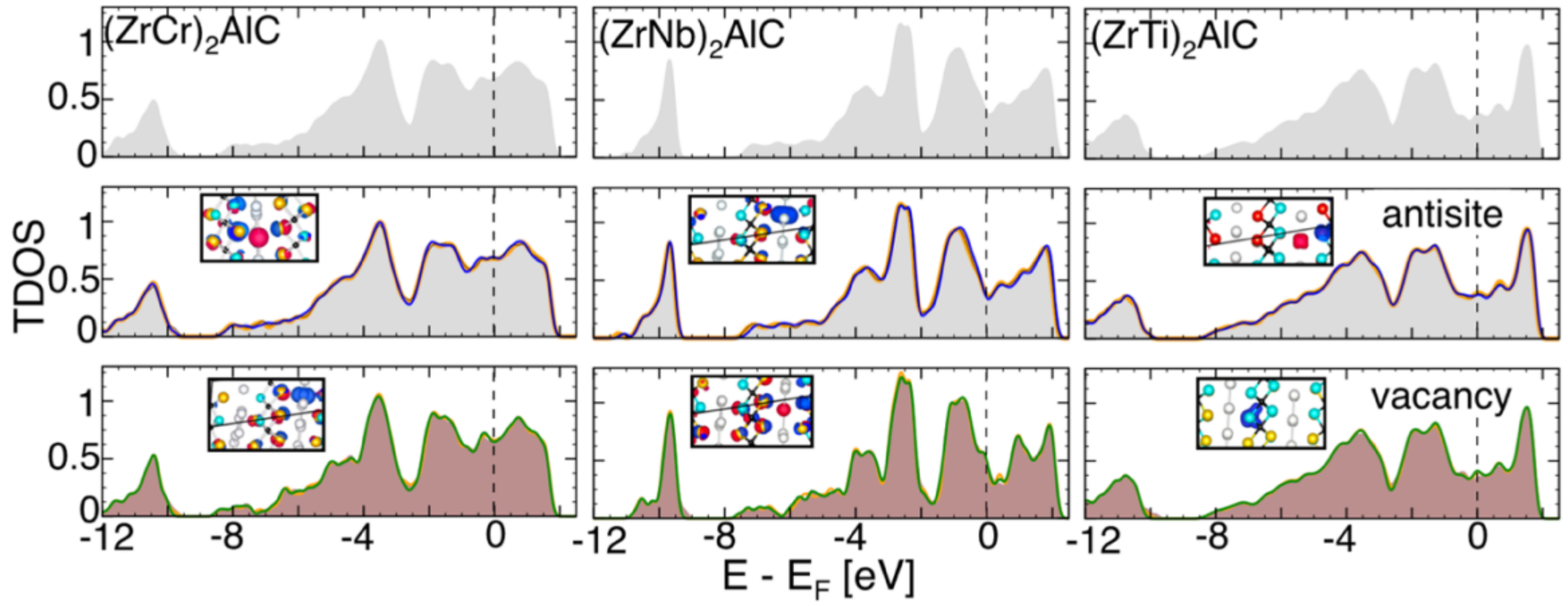

**Figure 5**. The total DOS of the pure (top), antisite (middle) disorder and monovacancy (bottom) defects in (M-M')2AlC (M=Zr; M'=Cr,Nb,Ti).

We note on comparing TDOS of $(Zr\text{-}M)_2AlC$ in **Fig. 5** with antisite (middle) and vacancy (bottom) defects that the hybridization energy peaks near the $E_{Fermi}$ for vacancy cases changes greatly. This also depends on type of vacancy, i.e., $V_M$, $V_A$, or $V_X$, which alters the electronic density and the strength of hybridization near $E_{Fermi}$ because $V_M$ and $V_C$ reduces the number of M-C bonds. On the contrary, for the A vacancy, the neighboring atoms readjusts itself to maintain the energy stability by charge transfer. Comparing total DOS in the presence of the vacancy (bottom-panel) and no vacancies (top-panel), we could clearly see the hybridization peak in the same energy range near the $E_{Fermi}$ for vacancy cases. We attribute this to the reduced numbers of M-A bonds. Notably, no such changes are observed for the case of antisite defects as system's electronic density (only atomic positions are interchanged) remains conserved

To further investigate this, we plot the charge density difference between vacancy and no-vacancy cases in the inset **Fig. 5**. The DOS above the $E_{Fermi}$ is the result of weaker Zr-*d*/Cr-*d* and Al-*p* bonding with a little contribution form the C-*p* states (also see supplementary material). On the other hand, the top of the valance band is formed by Zr-*d*/Cr-*d*, C-*p* and Al-*p* *hybridized* states. For the ordered MAX phase, in **Fig. S8**, the states at the $E_{Fermi}$ are mainly from the Zr-*d*/Cr-*d*, C-*p* and Al-*p* with a small contribution of C-*p*.

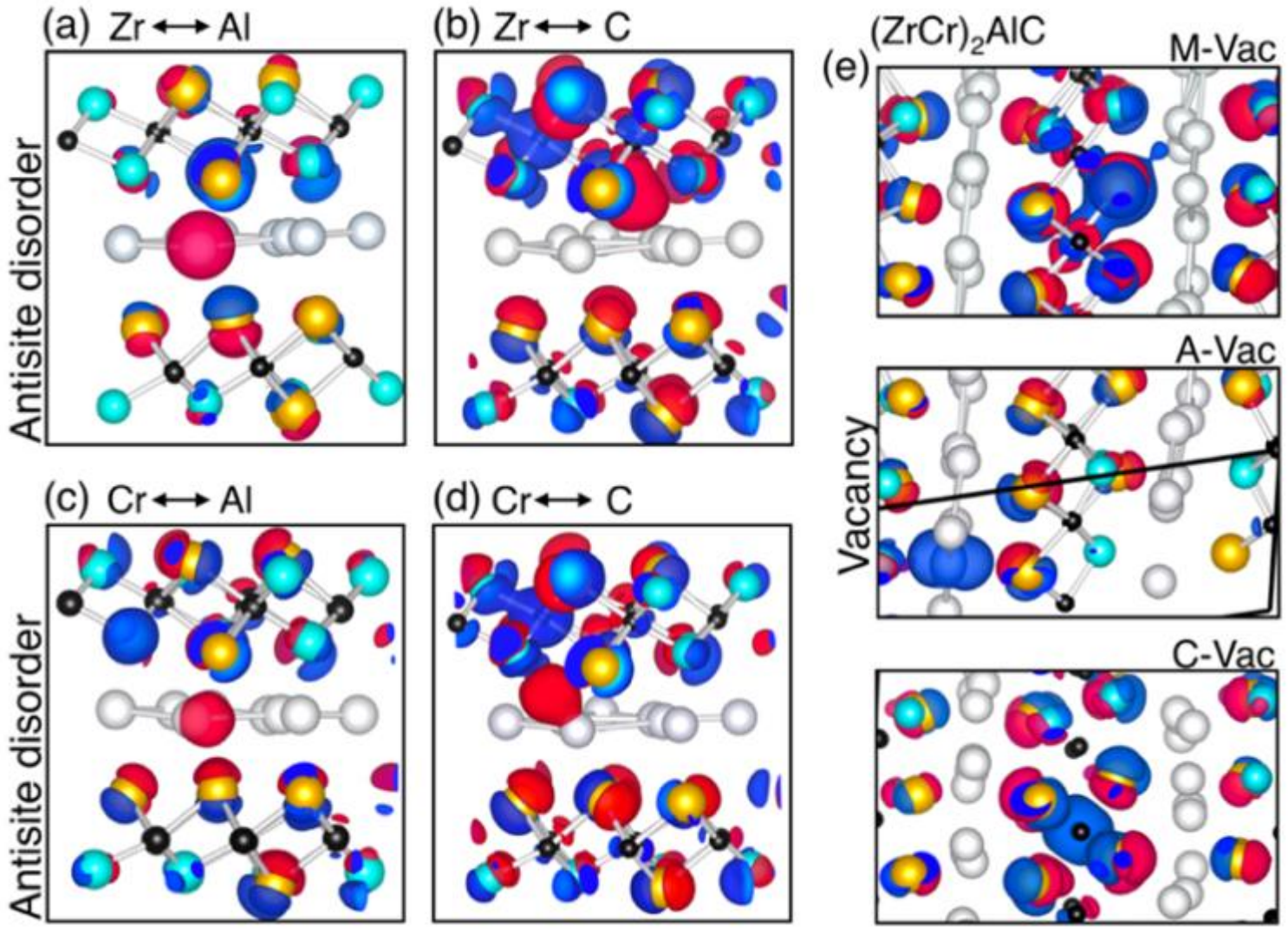

**Figure 6**. Charge difference of M-site disorder $(ZrCr)_2AlC$ MAX phase with (a-d) antisite disorder and (e) $V_M$, $V_A$ and $V_x$ vacancies. Clearly, antisite and vacancy defect changes the local chemical activity by stronger charge transfer.

**Charge density difference of antisite and vacancy defects with disorder $(ZrCr)_2AlC$**: The MAX phases form M-X, M-A, and A-A bonds, where M-X is the covalent and A-A is the metallic respectively. This is the reason M-X bonds are stronger compared to A-A. If we compare two ordered MAX phases, e.g., $Zr_2AlC$ and $Cr_2AlC$, the Zr-based compounds are weaker than the Cr-based compounds due to significant difference in $\chi$ between Zr and Cr ($\chi_{Zr}$ = 1.33, $\chi_{Cr}$= 1.66). Large $\chi$ of Cr attracts more charges and makes the bonds in $Cr_2AlC$ stronger. This also suggests that $Zr_2AlC$ will have more agility towards re-organizing defects compared to Cr-based MAX phases. The local bonds near the vacancy sites are disturbed due to vacancy formation as shown in **Fig. 6** by charge density difference between pure and defect $(ZrCr)_2AlC$. For example, the cationic defects lead to much stronger bonding between M(=Cr/Zr), viz. Zr-Zr/Cr-Cr and Al-C.

Therefore, we believe that alloying and defect (antisite/vacancy) can be used as an effective approach in modifying materials properties. The smaller atomic radius of Cr (0.76 Å) compared to Zr (0.86 Å) increases the C-Cr interaction compared C-Zr, which redistributed the electronic-density in and around the A-layer. We believe that this electronic rearrangement plays an important role in the increased stability of $(ZCr)_2AlC$. Electronegativity is another factor important to understand the chemical bonding. Following the Pauling electronegativity scale ($\chi_{Cr}$ =1.66, $\chi_{Zr}$ =1.33, $\chi_{Al}$ = 1.61 and $\chi_C$ = 2.55), we find a significant charge transfer in **Fig. 6** from Zr/Cr to C, where the Zr/Cr-C bonding exhibits directionality and therefore the covalent character. The driving force behind the displacement of the bonding charge is the greater ability of C to attract electrons as a result of the difference between atomic electronegativities ($\Delta\chi = \chi_M - \chi_C$).

**Vacancy migrations energies**: Vacancies are the most common defects at equilibrium and they greatly control the kinetics of diffusion-mediated processes. For high-temperature applications, vacancy diffusion is an important aspect to explore, particularly because the exchange between vacancies and atoms in the lattice is often times the limiting step when considering possible chemistry-driven transformations, including the selective oxidation of Al in Al-containing MAX phases, for example. To date, very little progress has been made towards increasing the understanding of vacancy diffusion mechanism in disordered MAX phases from a theoretical framework—we note that a companion paper from the present group explicitly looks at migration energies of vacancies for a large number of pure 211 MAX phase systems [49].

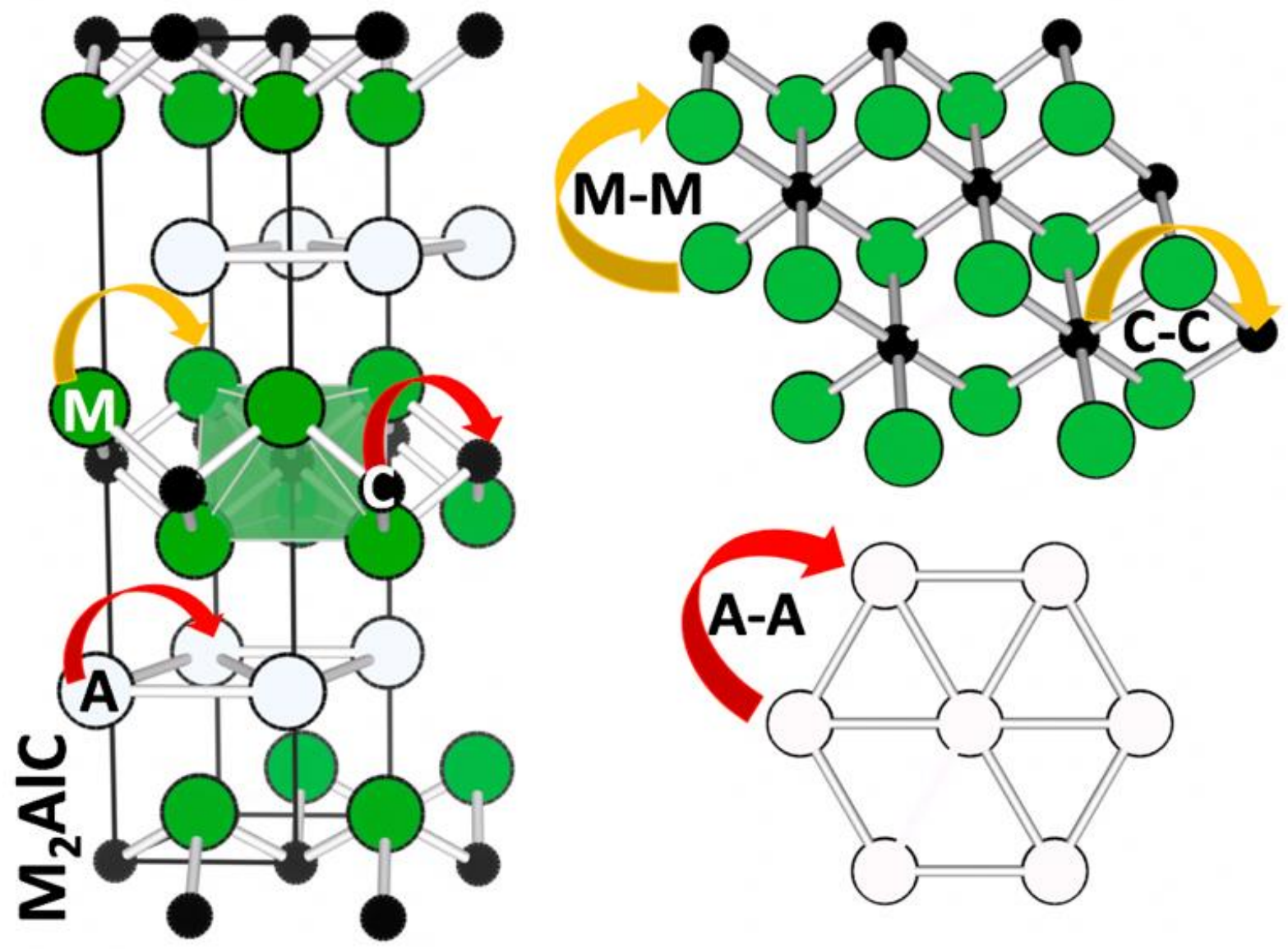

**Figure 7.** Schematic of $V_M$, $V_A$ and $V_C$ vacancy migration in order and disorder MAX phase alloy.

We study $V_M$, $V_A$, and $V_X$ migration in disordered MAX phases as shown by schematic in **Fig. 7**. We use cNEB to generate a succession of configurations (images) along the initial band connecting the initial-to-final state and calculate vacancy formation energies. cNEB relaxes each state on the band and provides information for deforming that band towards a lower energy embedding in the potential energy landscape. The band location in the energy landscape after sufficient number of iterations corresponds to the minimum path connecting the reactants and products.

In **Fig. 8(a-c)**, we plot the $V_M$, $V_{Al}$, and $V_C$ migration energies for five ordered MAX phase compounds: $M_2AlC$, M=Ti, Ta, Cr, Nb, Zr. The $Zr_2AlC$ shows lowest energy barrier for $V_M$, $V_{Al}$, and $V_C$ in comparison to all other case with samaller $V_{Al}$ diffusion barrier height. The overall order of vacancy migration is $V_M > V_{Al} > V_C$. The $Zr_2AlC$ is the only exception where $V_M$ has lower migration energy than $V_C$. Comparing $E_{form}^{Vac}$ in **Fig. 1b** with barrier energy in **Fig. 8a-c** of ordered MAX phase, we can make the inference that stable $V_{Al}$ is easier to move within the basal plane than other vacancies. In spite of competing migration energies of $V_M$ and $V_C$ in $Zr_2AlC$, $E_{form}^{Vac}$ is in sharp contrast in term of energy stability in **Fig. 1b**. Higher $E_{form}^{Vac}$ for $Zr_2AlC$ suggest that vacancies are hard to create, while (relatively) low diffusion barriers suggest facile vacancy migration.

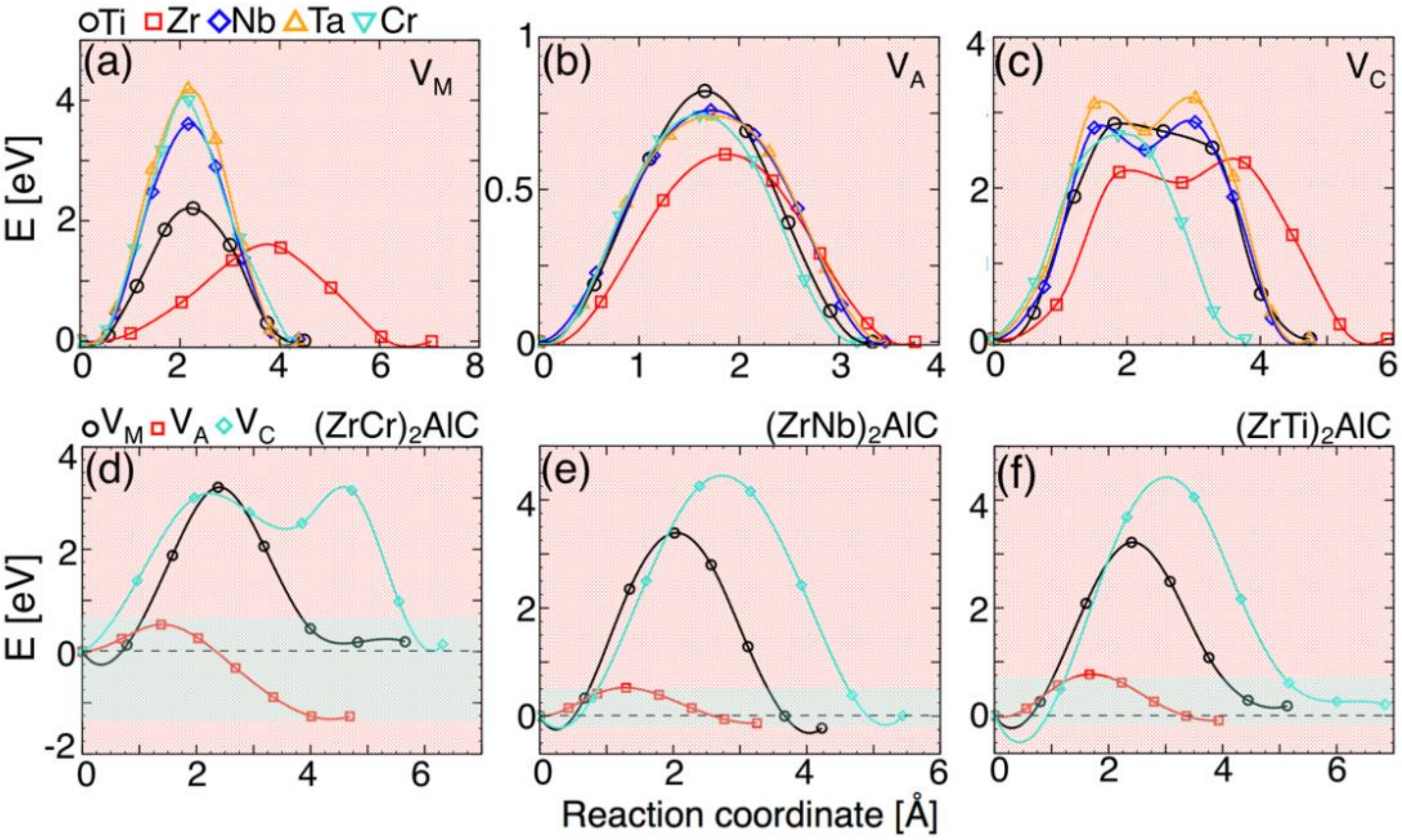

**Figure 8**. Calculated energy barrier of $V_M$, $V_A$ and $V_C$ vacancy migration energies in (a-c) ordered and (d-f) disorder MAX phases using cNEB method. We consider M-site disorder $(MM')_2AlC$, M, M'=Ti, Zr, Nb, Ta, Cr. The vacancy at A-sublattice, $V_A$, shows lower vacancy diffusion barrier compared to $V_M$ and $V_C$. The plot describes the energy variation along the minimum path. The energy threshold (y-axis), i.e., the activation energy, must be overcome for vacancy migration.

The vacancy migration barrier for disordered $(ZrM)_2AlC$ in **Fig. 8(d-f)** follows the trend of $E_{form}^{Vac}$ in **Fig. 3**, where A-site migration is easier in $(ZrNb)_2AlC$ compared to $(ZrCr)_2AlC$ and $(ZrTi)_2AlC$, i.e., $E_{mig}$ [$(ZrCr)_2AlC$] > $E_{mig}$ [$(ZrTi)_2AlC$] > $E_{mig}$ [$(ZrNb)_2AlC$]. The barrier energies in **Fig. 8(d-f)** for the disordered alloys seem to depend greatly on the local environment. This is further established by zero end-point energies in ordered MAX phases, whereas the non-zero end-point energies in disorder MAX phases are configuration dependent. Creating two vacancies at symmetrically related Wycoff-positions changes the neighbor distributions, which affects the total energy of the vacancy supercell. The vacancy formation energies and vacancy concentration can directly be connected. If we compare vacancy formation energies from **Fig. 2**, e.g., the lower and comparable formation energies of $(ZrNb)_2AlC$ and $(ZrTi)_2AlC$ will allow larger vacancy concentration compared to $(ZrCr)_2AlC$. Similarly, for $(ZrM)_2(AA')C$, the $V_A$ vacancy migration in $(ZrM)_2AlC$ within the basal plane is easier and proceeds with an energy barrier of 0.5 to 1.0 eV. This is much lower than the $V_M$ or $V_C$ vacancy migration barrier. However, the high-migration energy makes $V_M$ diffusion unfavorable compared to $V_A$ and $V_X$. Therefore, the significant population of the vacancies created in M-layers may remain intact even after a significant long time.

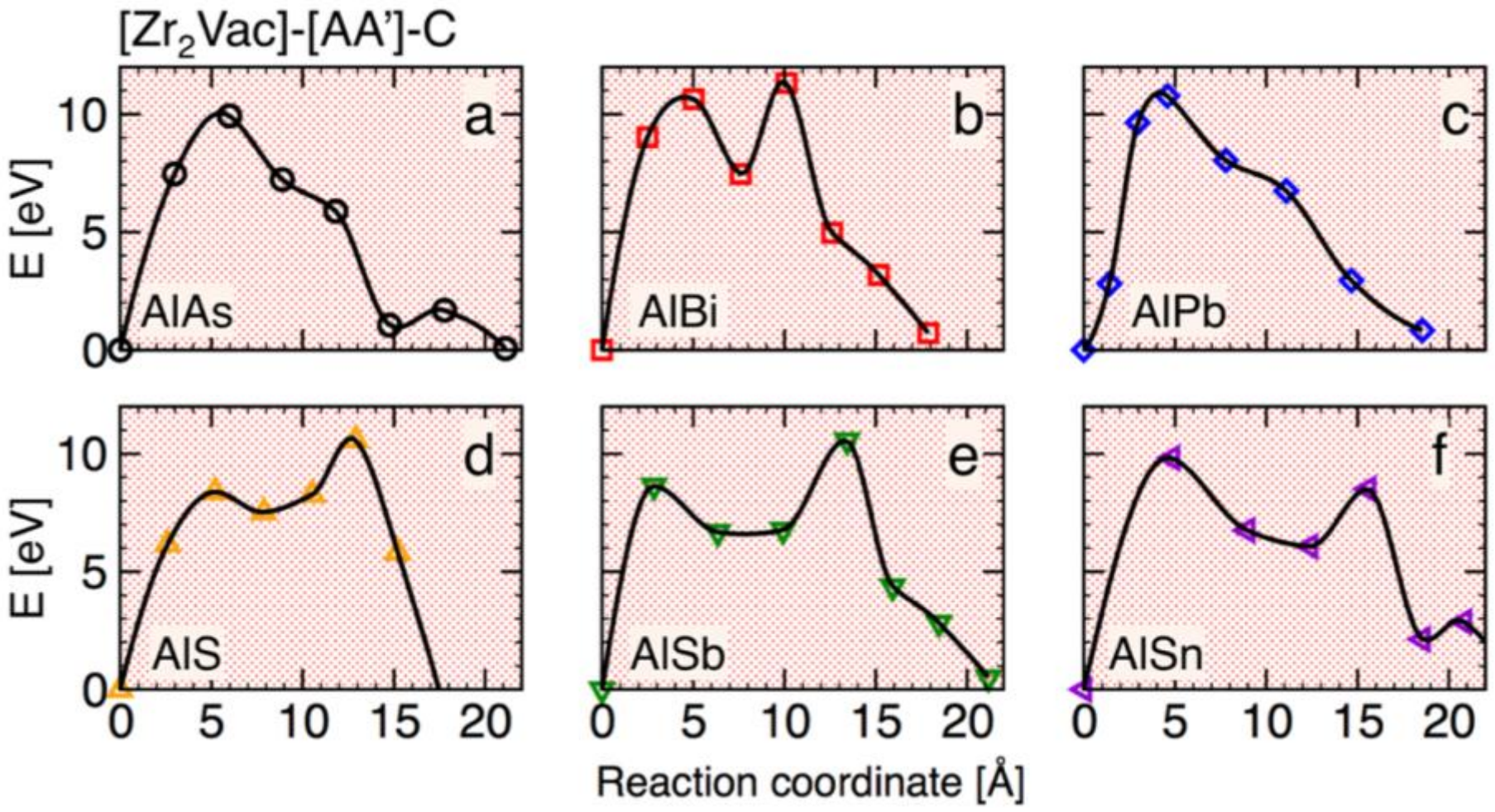

**Figure 9**. (a-f) The M-site ($V_M$) vacancy migration barrier for A-site disorder for $Zr_2$(A-A')C, A=Al, A'=As, Bi, Pb, S, Sb, Sn. The energy threshold (activation energy) must be overcome for vacancy migration. With A-site disorder, vacancy migration at Zr-site becomes harder.

In **Fig. 9** and **Fig. 10**, we present barrier energy for A-site disorder [$Zr_2$(A-A')C, A=Al, A'= As, Bi, Pb, S, Sb, Sn] and two (M and A)-site disorder ((Zr-M)$_2$(A-A')C, where M=Cr,Nb,Ti and A-A'=SnAl), respectively. The high vacancy migration energies in **Fig. 9** shows very high barrier of 9-12 eV for A-site disorder with $V_M$. The high barrier makes vacancy diffusion almost impossible, however, once the vacancies are created, the significant population of the vacancies will be intact for a considerable time under the irradiation process. On the other hand, for two-site disorder with $V_M$, $V_A$ and $V_C$ vacancies in **Fig. 10** (also see **Fig. S6**), the diffusion barrier show competing energies for $V_M$ and $V_C$ for M=Nb and Ti. The A-site vacancies for M=Cr show almost barrierless diffusion. If compared to single M-site disorder migration energies as presented in **Fig. 8d-f**, adding more disorder, i.e., to both to M- and A-site, further improves the vacancy diffusion by reducing the barrier height. The $V_M$ and $V_X$ site vacancies show competing $E_{form}^{Vac}$ in **Fig. 3b**, but vacancy migration becomes easier for $V_A$ with small barrier size of 0.40 eV due to increased chemical disorder.

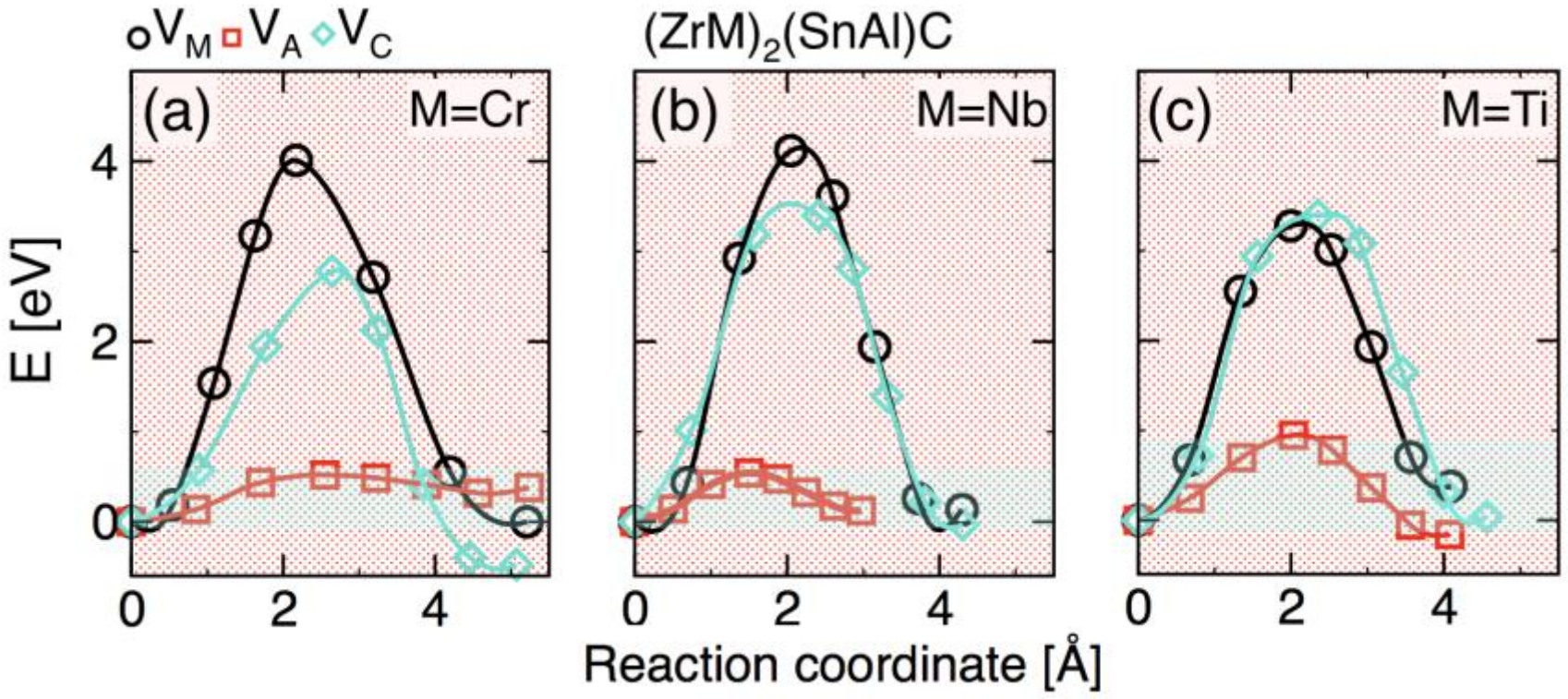

**Figure 10.** The $V_M$, $V_A$ and $V_C$ vacancy migration energies of MAX alloys with M- and A-site disorder using cNEB method: (M-M')$_2$(A-A')C, M=Zr, M'=Cr,Nb,Ti, A=Sn and A'=Al.

## Conclusion

MAX phases are very exciting new class of materials because of their high-temperature applications and the vacancies are most common defects at high operating conditions. Therefore, we set out the main objective of this work to find out: (1) How the role of vacancies changes in going from order to disorder MAX phases; and (2) Can we use alloying in combination with vacancies as an effective tool to tailor MAX phase properties? We perform first-principles DFT calculations to explore the alloying effect on the structural-stability, energy-stability, electronic-structure, and vacancy diffusion of MAX phases. Both as a test-bed and useful application in nuclear engineering, $Zr_2AlC$ was an interesting example to be explored for the effect of alloying and defect formation because the lower formation energy of ordered $Zr_2AlC$ suggests the possibility of higher vacancy concentration which may in turn be beneficial for the production of protective passive oxide layers.

We considered single-site (M-/A-) and two-site (M-A) disorder and explored the vacancy and antisite defect stability. We found that disordering has a significant effect on energy-stability, electronic-structure and vacancy migration behavior of $(Zr\text{-}M')_2(A\text{-}A')X$ (M = Cr, Nb, Ti, AA' = Al, SnAl, X = C), whereas in some cases antisite defects are much easier to form compared to point vacancies. For example, Al preferentially goes to Cr compared to Zr in Cr-doped $Zr_2AlC$, i.e., $(ZrCr)_2AlC$. Other focus of this work was to understand the effect of chemical alloying and defects on vacancy diffusion profile of MAX phases. While the diffusion barrier in strongly bonded metal alloys is usually large, our DFT+cNEB calculations show that alloying helps in lowering down both vacancy formation energy and diffusion-barrier compared to ordered MAX phases. The reduced barrier height for Al-diffusion as found in our study through alloying is an important finding because Al is known to promote the formation of protective oxide layer (i.e., $Al_2O_3$ in aluminum-based MAX phases) at high-operating temperatures. Therefore, the reduced vacancy formation energy and barrier height will help the formation of protective oxide layer in Al-based MAX phases. Based on our study, we believe that the chemical alloying route will surely help the MAX phase community to understand the defect formation and migration mechanism as well as provide ways to manipulate the electronic and mechanical behavior of MAX phases.

## Acknowledgement

Valuable discussion with Dr. M. Radovic is acknowledged. The financial support was provided from National Science Foundation through grants no. (DMREF) CMMI-1729350. First-principles calculations were carried out at the Supercomputing Facility at Texas A&M University. DS acknowledges the support of NSF through grant no. NSF-DGE-1545403.

*prashant40179@gmail.com

Supplemental material

# ***The Effect of Chemical Disorder on Defect Formation and Migration in Disordered MAX Phases***

Prashant Singh,[a] Daniel Sauceda,[a] and Raymundo Arroyave[a]

[a]*Department of Materials Science & Engineering, Texas A&M University, College Station, TX 77843, USA*

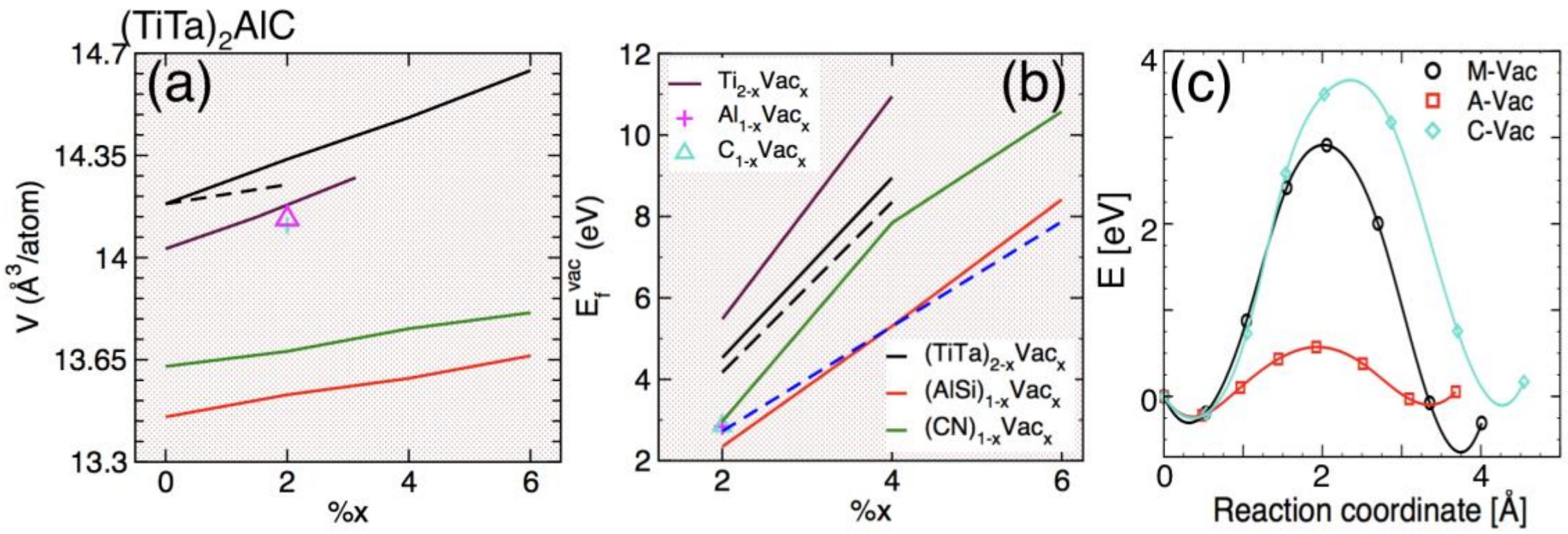

**Figure S1**. Vacancy dependence change in (a) volume, (b) formation energy in $(TiTa)_2AlC$, $Ti_2(AlSi)C$, and TiAl(CN). (c) Energy barrier of $V_M$, $V_A$ and $V_C$ vacancy migration in $(TiTa)_2AlC$. We calculate vacancy migration minimum energy path at the equilibrium volume in a 200-lattice supercell.

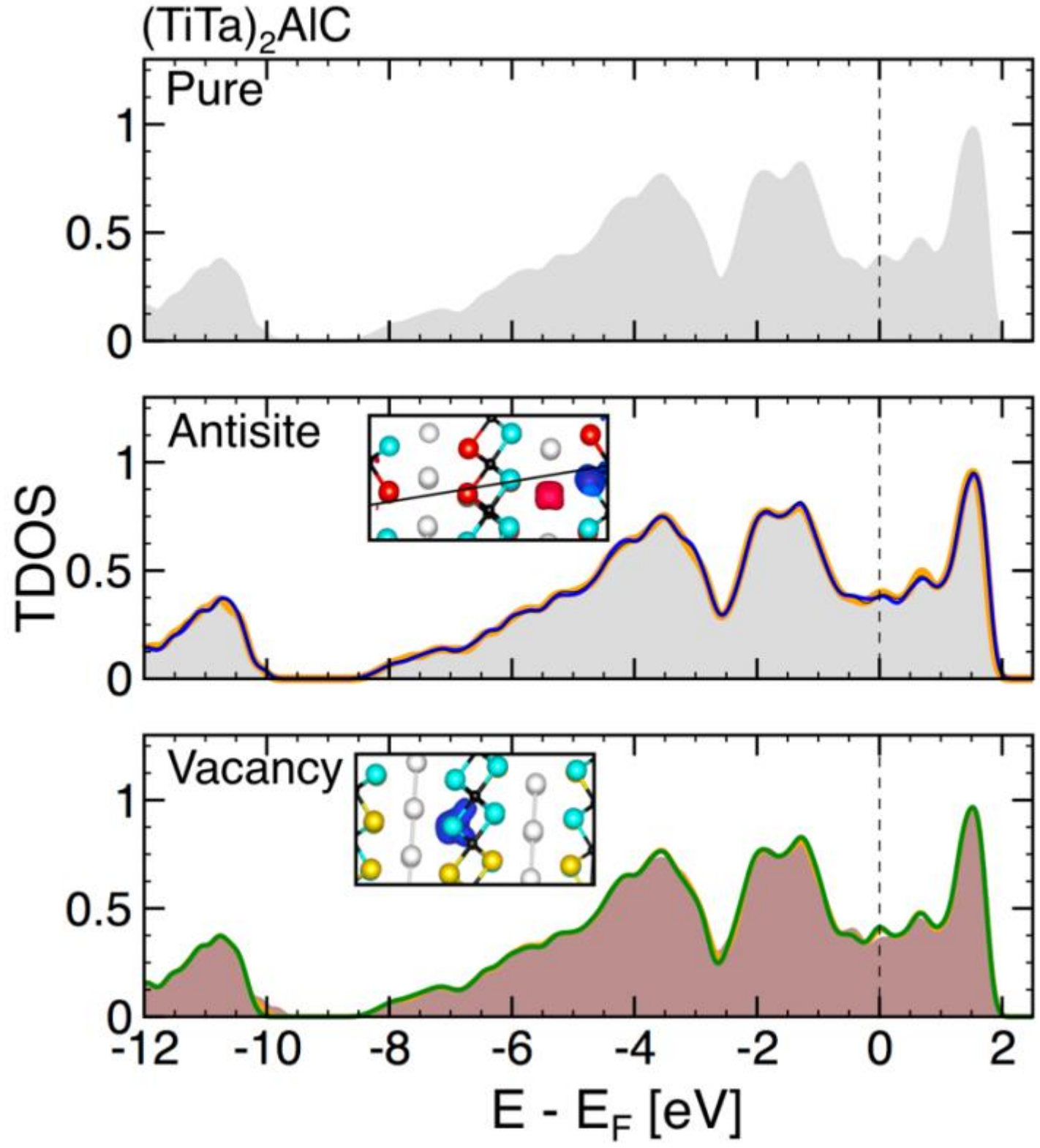

**Figure S2**. The total DOS of the pure (top), antisite (middle) disorder and monovacancy (bottom) defects in $(TiTa)_2AlC$.

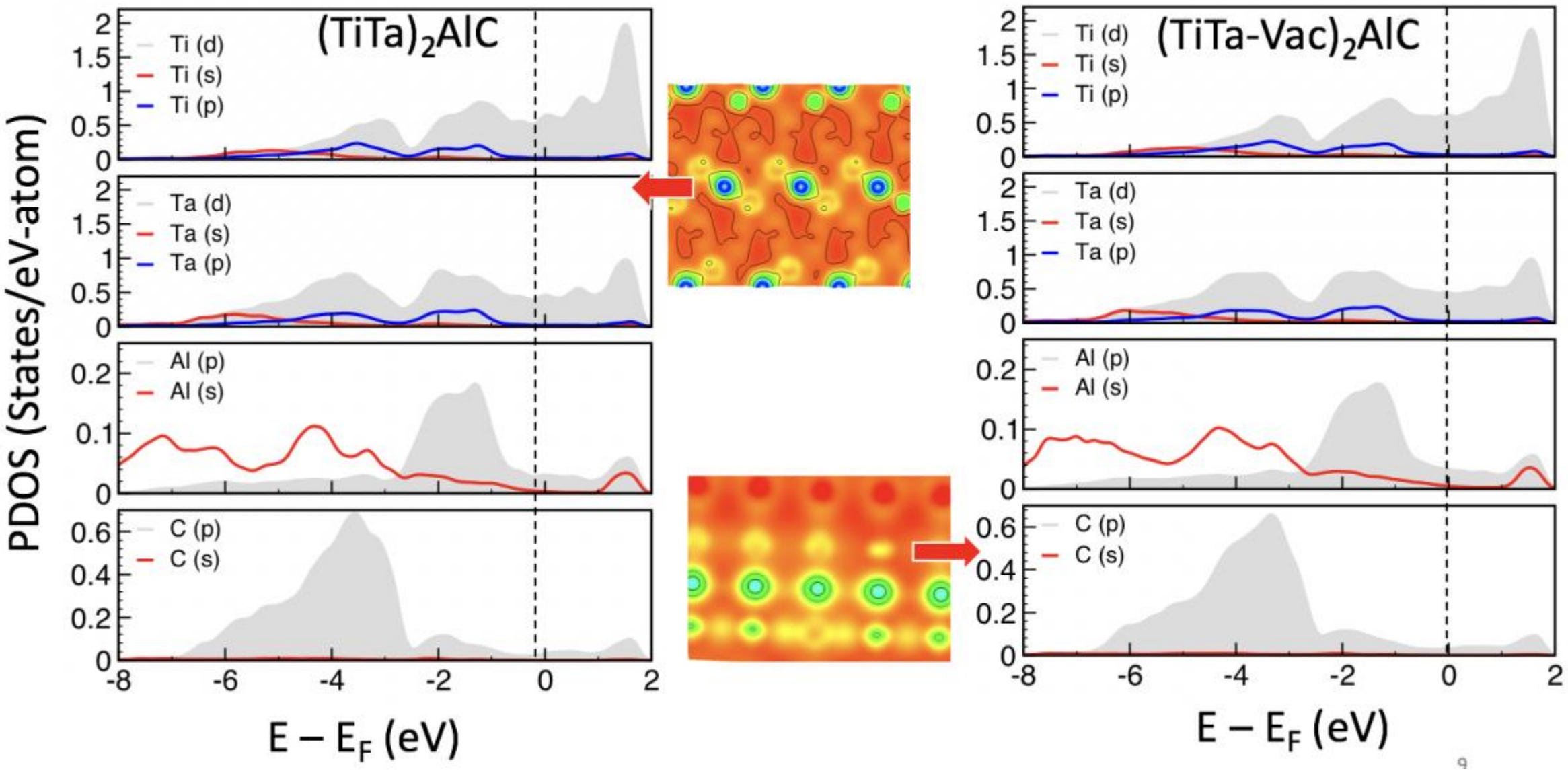

**Figure S3**. The partial DOS of the pure (left-panel) and vacancy (right-panel) in disorder $(TiTa)_2AlC$. 2D charge density plot showing charge depletion at the vacancy site.

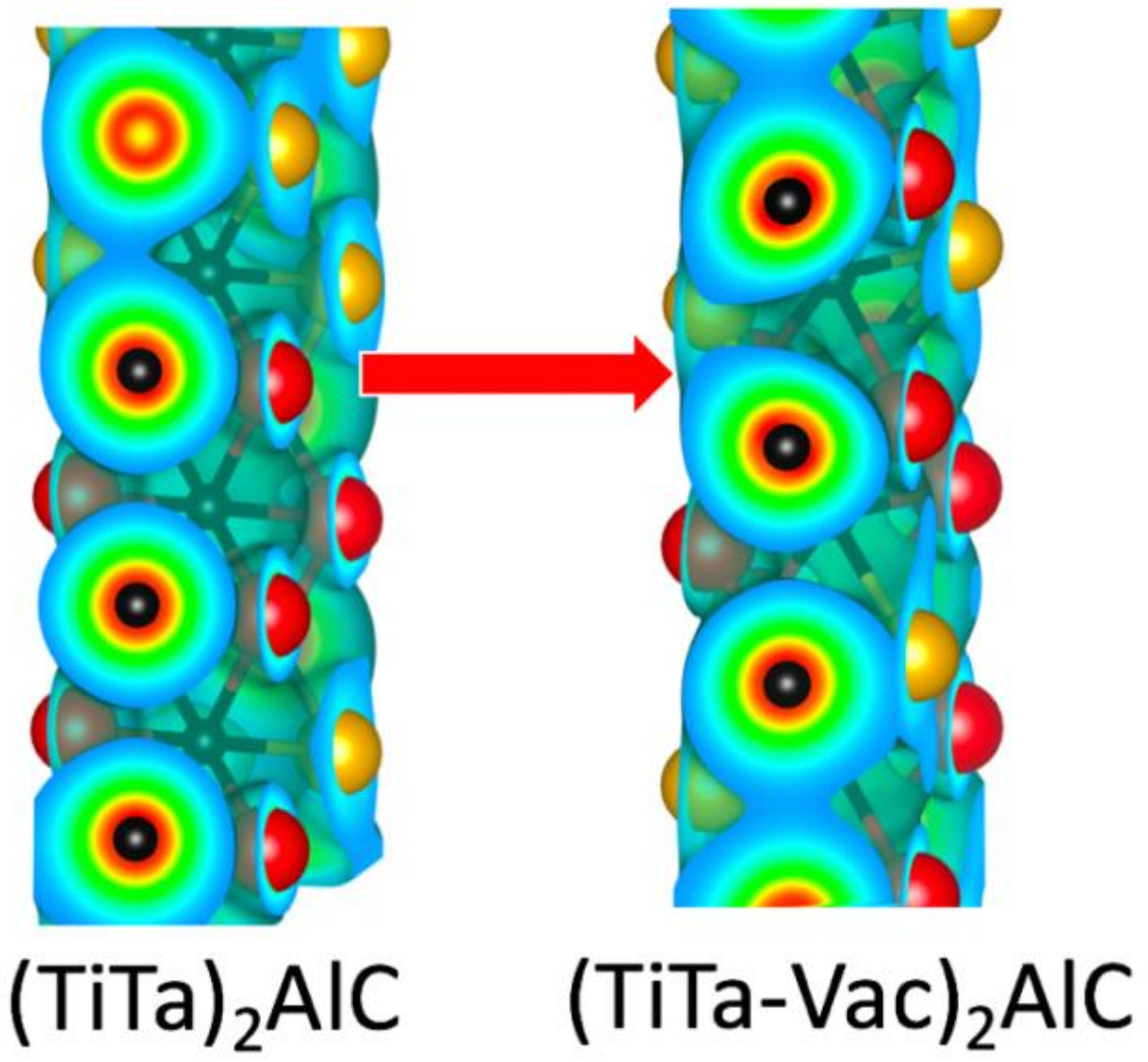

**Figure S4**. The partial charge-densities of the pure (left-panel) and vacancy (right-panel) in disorder $(TiTa)_2AlC$.

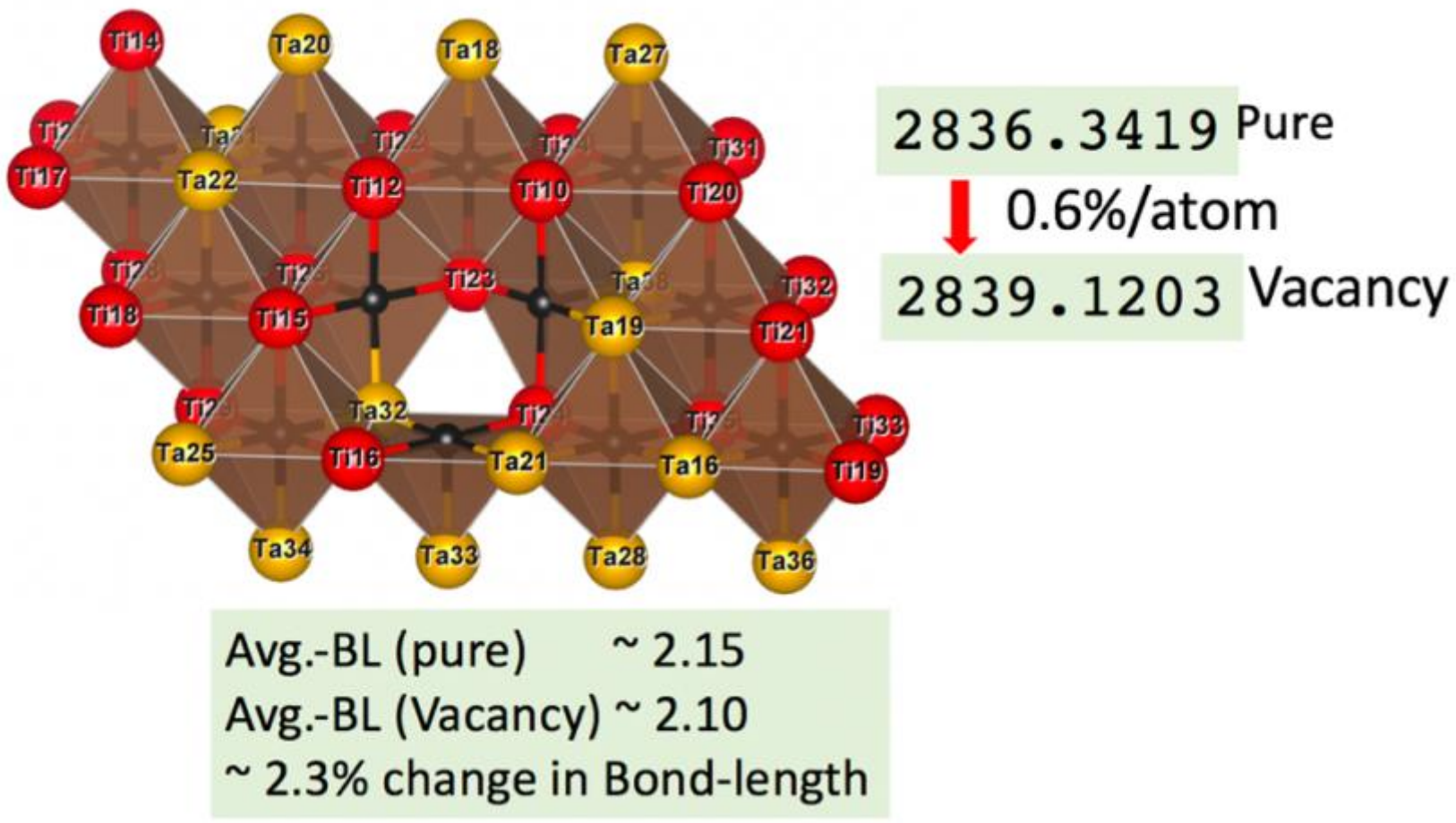

**Figure S5**. The volume and average bond-length change of +0.6% and 2.3% for M-site disorder with M-vacancy with respect to $(TiTa)_2AlC$ with no vacancies.

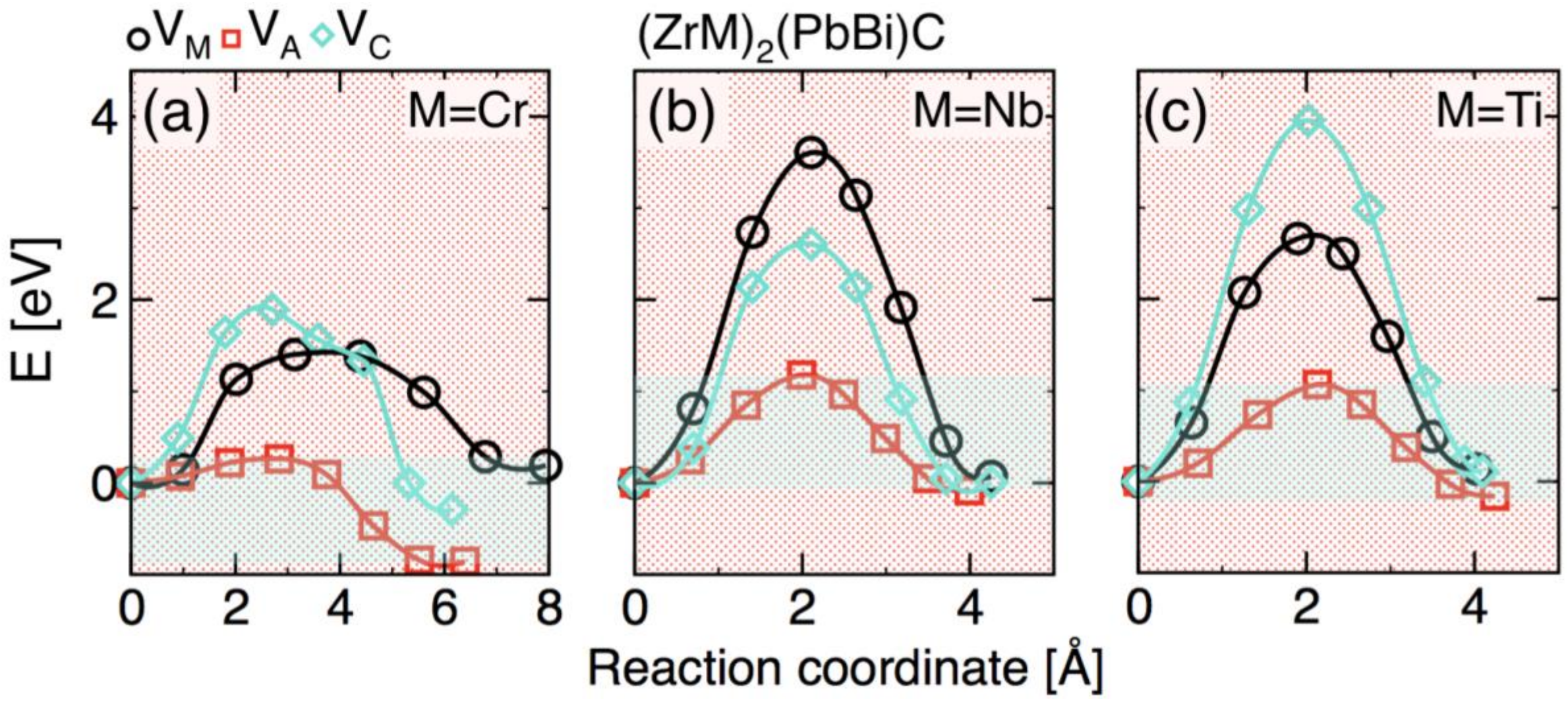

**Figure S6**. The $V_M$, $V_A$ and $V_C$ vacancy migration energies of MAX alloys with M- and A-site disorder using cNEB method: $(MM')_2(AA')C$, M=Zr, M'=Cr,Nb,Ti, A=Sn and A'=Al. We use 200 atom supercell to calculate minimum energy path for vacancy migration at the equilibrium volume.

***Electronic structure***: The local feature of the TDOS near the Fermi-level ($E_{Fermi}$) is indicative of the stability. For example, a valley in DOS at $E_{Fermi}$ signifies a lower energy barrier for electrons (E < 0 eV) to move to the unoccupied states, whereas peak usually implies higher energy barrier (E>0 eV). This criterion works reasonably well for **Fig. S7 & S8**. $Ti_2AlC$ and $Ta_2AlC$ have a valley at $E_{Fermi}$, suggests increased stability. Out of 9-MAX phase ordered alloys in **Fig. S7**, only $Zr_2AlC$ shows a peak at $E_{Fermi}$ in the TDOS. Our qualitative assumption explains the observation of the early transition metal based MAX phases very well [S**1**].

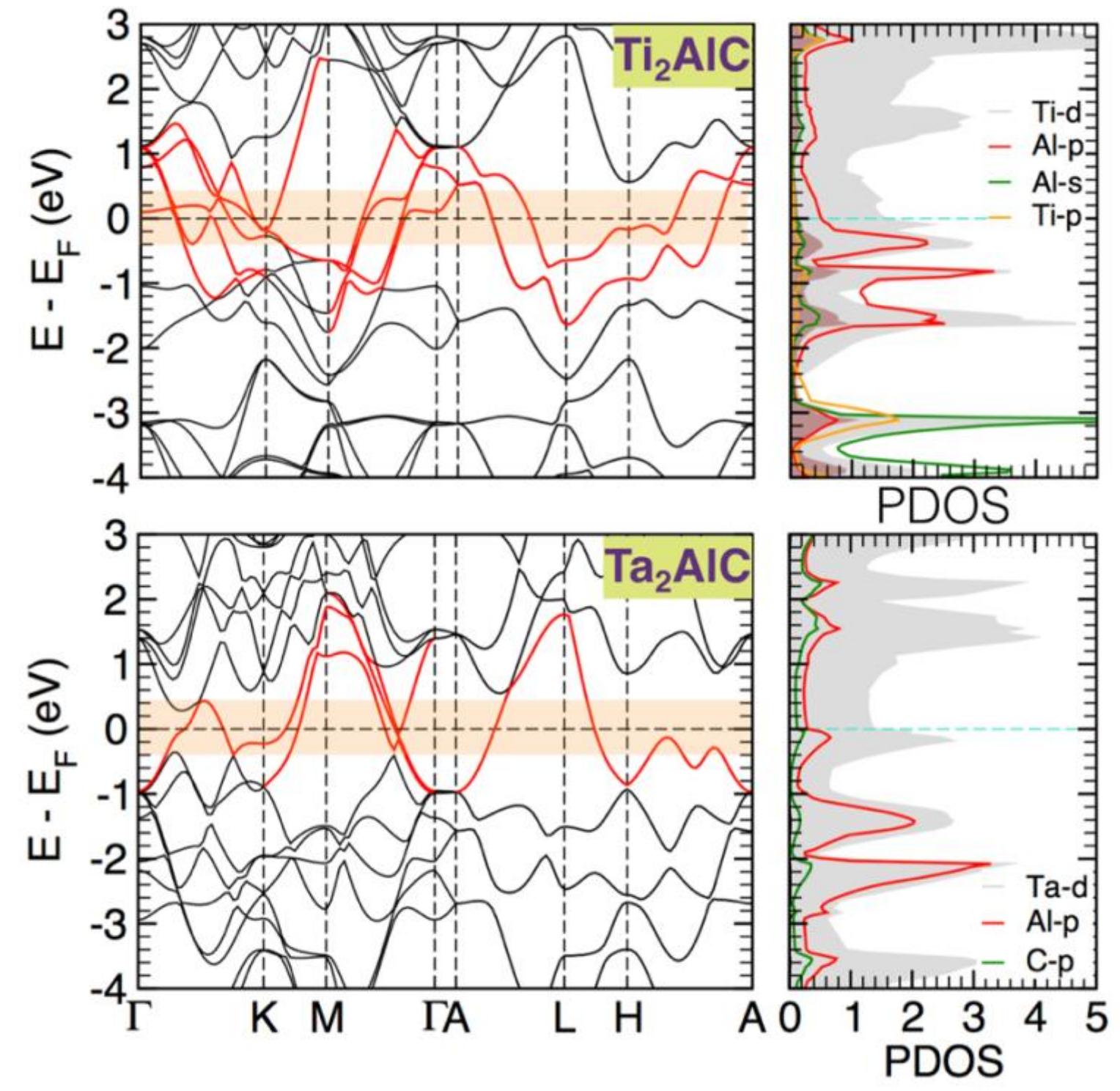

**Figure S7**. Electronic band-structure and partial density of states of $Ti_2AlC$ and $Ta_2AlC$ ordered MAX phase alloys.

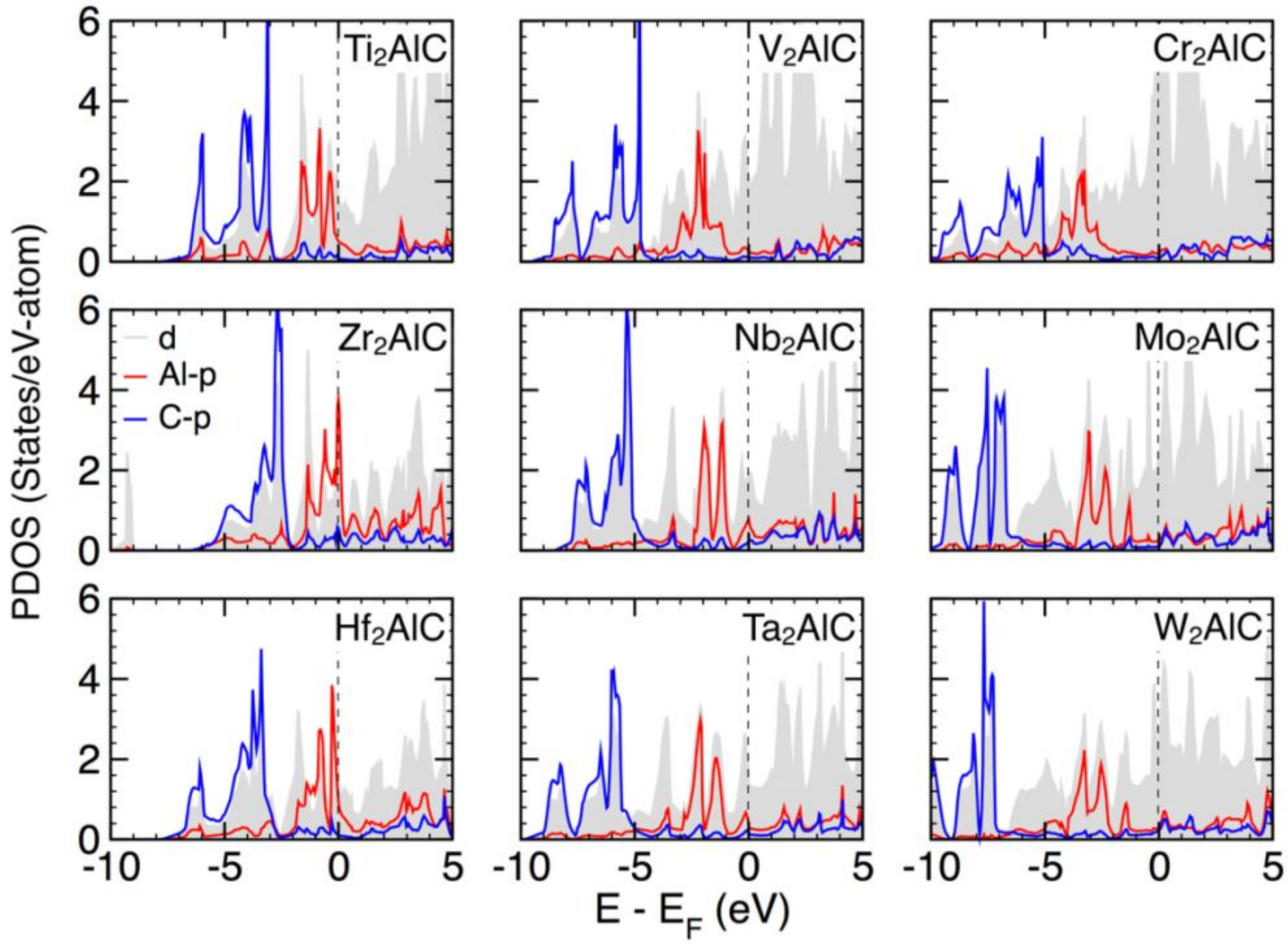

**Figure S8**. Partial density of states (PDOS) of $M_2AlC$ ordered MAX phase alloys for M=Ti, V,Cr,Zr,Nb,Mo,Hf,Ta,W. Only $Zr_2AlC$ shows a peak exactly on Fermi-level, which indicates alloy instability.

| System | a | b | c | V | $C_{11}$ | $C_{12}$ | $C_{44}$ | K | G | E | ν | k |
|---|---|---|---|---|---|---|---|---|---|---|---|---|
| | Å | | | Å3 | GPa | | | | | | | |
| $Cr_2AlC$ | 2.86 | 2.48 | 12.70 | 90.2 | 334.8 | 74.1 | 131.1 | 175.1 | 126.0 | 304.9 | 0.21 | 1.39 |
| $Ta_2AlC$ | 3.10 | 2.69 | 13.92 | 115.9 | 332.7 | 128.6 | 159.0 | 194.1 | 121.5 | 301.6 | 0.24 | 1.60 |
| $Nb_2AlC$ | 3.14 | 2.72 | 13.91 | 118.7 | 323.9 | 83.4 | 141.7 | 170.1 | 119.6 | 290.7 | 0.22 | 1.42 |
| $Ti_2AlC$ | 3.05 | 2.64 | 13.78 | 111.2 | 293.1 | 65.2 | 107.7 | 139.2 | 107.5 | 256.6 | 0.19 | 1.29 |
| $Zr_2AlC$ | 3.31 | 2.87 | 14.63 | 138.9 | 253.5 | 72.2 | 90.5 | 130.1 | 87.6 | 214.7 | 0.22 | 1.48 |
| $Zr_2AlC$-$V_M$ | 6.65 | 5.76 | 14.68 | 561.5 | 214.5 | 70.9 | 51.4 | 112.8 | 63.2 | 159.9 | 0.26 | 1.78 |
| $Ti_2AlC$-$V_M$ | 6.12 | 5.30 | 13.81 | 448.3 | 257.4 | 73.3 | 75.8 | 125.4 | 85.2 | 208.5 | 0.22 | 1.47 |
| $Ta_2AlC$-$V_M$ | 6.18 | 5.35 | 13.90 | 460.1 | 328.5 | 111.6 | 133.4 | 184.6 | 114.7 | 285.1 | 0.24 | 1.61 |
| $Nb_2AlC$-$V_M$ | 6.25 | 5.42 | 13.90 | 470.7 | 293.5 | 88.6 | 116.7 | 160.8 | 102.5 | 253.8 | 0.24 | 1.57 |
| $Cr_2AlC$-$V_M$ | 5.69 | 4.92 | 12.62 | 353.3 | 322.3 | 64.2 | 122.6 | 163.2 | 122.2 | 293.4 | 0.20 | 1.34 |
| $Zr_2AlC$-$V_X$ | 6.64 | 5.75 | 14.63 | 559.2 | 228.9 | 72.4 | 82.7 | 121.9 | 79.2 | 195.5 | 0.23 | 1.54 |
| $Ti_2AlC$-$V_X$ | 6.11 | 5.29 | 13.75 | 445.2 | 267.7 | 68.5 | 101.4 | 134.0 | 99.6 | 239.5 | 0.20 | 1.34 |
| $Ta_2AlC$-$V_X$ | 6.16 | 5.33 | 13.99 | 459.5 | 318.8 | 138.0 | 141.8 | 197.0 | 111.8 | 282.2 | 0.26 | 1.76 |
| $Nb_2AlC$-$V_X$ | 6.22 | 5.39 | 14.02 | 469.7 | 283.7 | 120.6 | 124.7 | 173.9 | 99.4 | 250.5 | 0.26 | 1.75 |
| $Cr_2AlC$-$V_X$ | 5.65 | 4.89 | 12.65 | 349.8 | 359.3 | 86.1 | 138.7 | 188.6 | 133.7 | 324.5 | 0.21 | 1.41 |
| $Zr_2AlC$-$V_A$ | 6.69 | 5.79 | 14.45 | 559.3 | 243.4 | 62.1 | 58.7 | 118.5 | 74.1 | 184.1 | 0.24 | 1.60 |
| $Ti_2AlC$-$V_A$ | 6.16 | 5.33 | 13.61 | 446.7 | 277.0 | 62.5 | 70.5 | 128.7 | 88.7 | 216.5 | 0.22 | 1.45 |
| $Ta_2AlC$-$V_A$ | 6.20 | 5.37 | 13.85 | 460.5 | 324.6 | 129.8 | 126.0 | 186.2 | 106.7 | 268.9 | 0.26 | 1.74 |
| $Nb_2AlC$-$V_A$ | 6.29 | 5.45 | 13.72 | 471.3 | 301.4 | 90.0 | 105.7 | 161.8 | 99.2 | 247.2 | 0.25 | 1.63 |
| $Cr_2AlC$-$V_A$ | 5.69 | 4.92 | 12.62 | 353.3 | 332.3 | 67.4 | 121.4 | 166.9 | 123.3 | 297.0 | 0.20 | 1.35 |

**Table S1**- Structural properties of pure ordered MAX phases $M_2AlC$ with no-vacancy and with-vacancy ($V_M$, $V_A$, $V_X$).

| System | a | b | c | V | Zr-C | Cr-C | Al-Al |
|---|---|---|---|---|---|---|---|
| | Å | | | $Å^3$/atom | $Avg_{BL}$ (Å) | | |
| **No-vac** | 8.174 | 13.979 | 25.355 | 14.077 | 2.23 | 2.02 | 2.89 |
| $V_M$ | 8.142 | 14.002 | 25.330 | 14.118 | 2.24 | 1.96 | 2.78 |
| $V_A$ | 8.159 | 14.001 | 25.372 | 14.147 | 2.22 | 2.02 | 2.79 |
| $V_X$ | 8.158 | 14.000 | 25.317 | 14.136 | 2.27 | 1.98 | 2.96 |

**Table S2**- Structural properties of disorder $(ZrCr)_2AlC$ MAX phases with no-vacancy and $V_M$, $V_A$, and $V_X$ vacancies.